\documentclass[a4paper,fleqn,usenatbib]{mnras}

\usepackage[T1]{fontenc}
\usepackage{ae,aecompl}

\usepackage{graphicx}	
\usepackage{amsmath}	
\usepackage{amssymb}	
\usepackage{xfrac}
\usepackage{makecell}
\usepackage{ulem}

\newcommand{\kms}{km\,s$^{-1}$}	
\newcommand{\Msold}{M$_{\odot}$\,yr$^{-1}$}

\title[Morphology and kinematics of the gas envelope of L1527]{Morphology and kinematics of the gas envelope of protostar L1527 as obtained from ALMA observations of the C$^{18}$O(2-1) line emission}
\author[P. Tuan-Anh et al.]{
{P. Tuan-Anh\thanks{E-mail: ptanh@vnsc.org.vn}, P. T. Nhung, D. T. Hoai, P. N. Diep, N. T. Phuong, N. T. Thao}
\newauthor{ and P. Darriulat}
\\
Department of Astrophysics, Vietnam National Satellite Center, VAST, 18 Hoang Quoc Viet, Hanoi, Vietnam\\
}
\date{Accepted XXX. Received YYY; in original form ZZZ}

\pubyear{2016}

\begin{document}
\label{firstpage}
\pagerange{\pageref{firstpage}--\pageref{lastpage}}
\maketitle

\begin{abstract}
Using ALMA observations of the C$^{18}$O(2-1) line emission of the gas envelope of protostar L1527, we have reconstructed its morphology and kinematics under the assumption of axisymmetry about the west-east axis. The main original contribution to our understanding of the formation process of L1527 is the presentation of a simple 3D parameterisation based solely on regions that are not dominated by absorption. In the explored range ($\sim$0.7 to 5 arcsec from the star) the model reproduces observations better than earlier attempts. The main results include: a measurement of the rotation velocity that confirms its evolution to Keplerian toward short distances; a measurement of the mean in-fall velocity, 0.43$\pm$0.10 \kms, lower than free fall velocity, with no evidence for the significant $r$-dependence suggested by an earlier analysis; a measurement of the central mass, 0.23$\pm$0.06 M$_{\odot}$ within a distance of 1.5 arcsec from the star, in agreement with earlier estimates obtained from a different range of distances; evidence for a strong disc plane depression of the in-falling flux resulting in an $X$ shaped flow possibly caused by the freeze-out of CO molecules on dust grains; a measurement of the accretion rate, 3.5$\pm$1.0 10$^{-7}$\Msold at a distance of 1 arcsec (140 au) from the star; evidence for a 10$^\circ$ tilt of the symmetry plane of the envelope about the line of sight, cancelling below $\sim$3 arcsec from the star, but matching infrared observations and being also apparent on the sky map of the mean Doppler velocity.
\end{abstract}




\section{Introduction}

L1527 IRS is a class 0 protostar, at a distance of $\sim$140 pc, in its earliest stages of formation. It has a mass of $\sim$0.2 solar masses and is surrounded by a flared rotating envelope of about one solar mass, as recently reported by \citet{Tobin2012, Tobin2013} who observed it in the continuum (at 1.3 and 3.4 mm wavelength) and on the $^{13}$CO(2-1) line using the Submillimeter Array (SMA) and the Combined Array for Research in Millimeter Astronomy (CARMA). These observations suggest a mass accretion rate of \mbox{$\sim$6.6 10$^{-7}$ \Msold} and a disc diameter at the arcsecond scale. They were followed by measurements and analyses of the 1.3 mm dust polarization by \citet{Segura-Cox2015} and \citet{Davidson2014}, which reveal a toroidal magnetic field configuration.

ALMA observations of molecular lines have been reported very recently: of C$^{18}$O and SO by \citet{Ohashi2014}, of cyclic-C$_3$H$_2$ and SO by \citet{Sakai2014a} and of CCH, CS, H$_2$CO and CH$_3$OH by \citet{Sakai2014b}. All above observations establish the presence of in-falling cool gas (T$\sim$30 K) onto a rotating disc surrounding the protostar and having its axis close to the east-west direction, here taken as $x$ axis. Surprisingly, they have also revealed major differences between the responses of SO, H$_2$CO and CH$_3$OH and those of CO, CCH, CS and cyclic-C$_3$H$_2$: the latter display double-peaked Doppler velocity distributions while the former display single peaks. This is interpreted by the authors as evidence for different molecules populating different radial regions, the former being confined to the inner part of the envelope, close to the disc, and absent from the foreground gas, while the latter are distributed over the foreground gas and subject to absorption. The in-falling motion of the foreground gas explains the small red-shift of the observed absorption dip. The presence of a Keplerian disc having a diameter at the arcsecond scale is generally accepted, the rotation velocity of the envelope decreasing faster at larger distances. The mechanism governing the disc dynamics, in particular the relative roles of magnetohydrodynamic turbulences and of self-gravitation are still unclear \citep[for a recent review see][]{Li2014}.

The above mentioned very recent observations follow a long series of earlier observations that have established the presence of an outflow along the east-west direction, with an opening angle of some 40$^\circ$ \citep{Tobin2008, Tobin2010} and tilted eastward by some 10$^\circ$ from north. We refer the reader to the abundant literature quoted in the recent publications in the reference list attached to the present article.
In what follows, we produce a simple model of the gas envelope that relies exclusively on its morphologic and kinematic properties, without reference to what is causing the strong absorption detected above systemic velocity. This is at variance with earlier models and contributes significant new information to the understanding of the underlying accretion mechanism.\\

\section{The data sample}
\subsection{Observations and data reduction}
The observations used in the present article were made in cycle 1 of ALMA operation on July 20, 2014. The time spent on source was 36 minutes. The number of antennas was 35 for part of the time, 36 otherwise, the longest baseline being 783.5 m; the two sets of data have been merged and reduced by the ALMA staff. The systemic velocity used as origin of the Doppler velocity scale is \mbox{5.84 {\kms}}. The maps of the C$^{18}$O(2-1) line emission are centred on the maximum of the continuum emission with right ascension of \mbox{{4\,h 35\,m 53.870\,s}} and declination of \mbox{26$^\circ$ 3 arcmin 9.54 arcsec}. The synthetized beam size (FWHM) is 0.6 $\times$ 0.5 arcsec$^2$ with a $-174^\circ$ position angle from north. These data have a better spatial resolution but are otherwise similar to the cycle 0 observations used by \citet{Ohashi2014}. Pixels are 0.15 $\times$ 0.15 arcsec$^2$ and the velocity bins are 0.16 {\kms} wide. The data used in the present work are  limited to a region in the plane of the sky covering 17 pixels in $x$ and 57 pixels in $y$, namely 2.55 and 8.55 arcsec respectively, over which a significant signal can be reliably detected. In practice we limit the  analysis to distances of the order of 4 arcsec where filtering out at large distances is not significant, the minimum baselines being essentially the same here and in the observations used by \citet{Ohashi2014}. We use orthonormal coordinates with $z$ along the line of sight, $x$ and $y$ in the plane of the sky, $x$ pointing east and $y$ pointing north. Throughout the article we assume a distance of 140 pc from the Sun, implying that an angular distance of 1 arcsec corresponds to an actual distance of 140 au in the plane of the sky. Accordingly, we use indifferently arcsec or au to measure distances, including along the line of sight.

As mentioned earlier, the continuum data are only used to define the origin of coordinates. Indeed, Gaussian fits to the $x$ and $y$ distributions of the continuum flux give full widths at half-maximum (FWHM) of 0.47 and 0.71 arcsec respectively with the same beam size as for the line data: the source is unresolved in $x$ and has a size at the scale of $\sim$0.4 arcsec in $y$, consistent with the size of the disc.
\subsection{Line data}
\begin{figure}
\centering
\includegraphics[height=7.5cm,trim=0.cm 0.cm 0.cm 0.cm,clip]{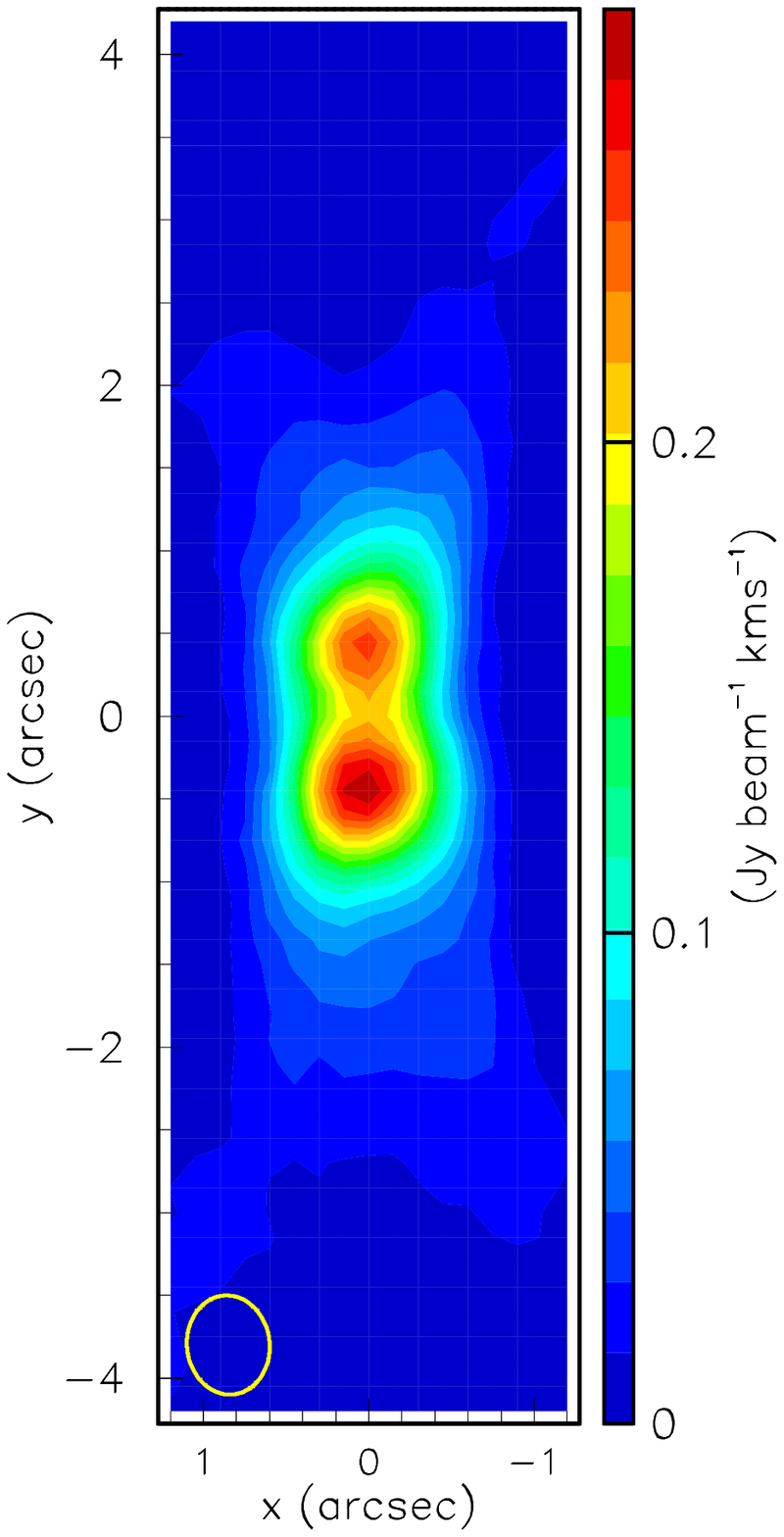}
\includegraphics[height=7.5cm,trim=0.cm 0.cm 0.cm 0.cm,clip]{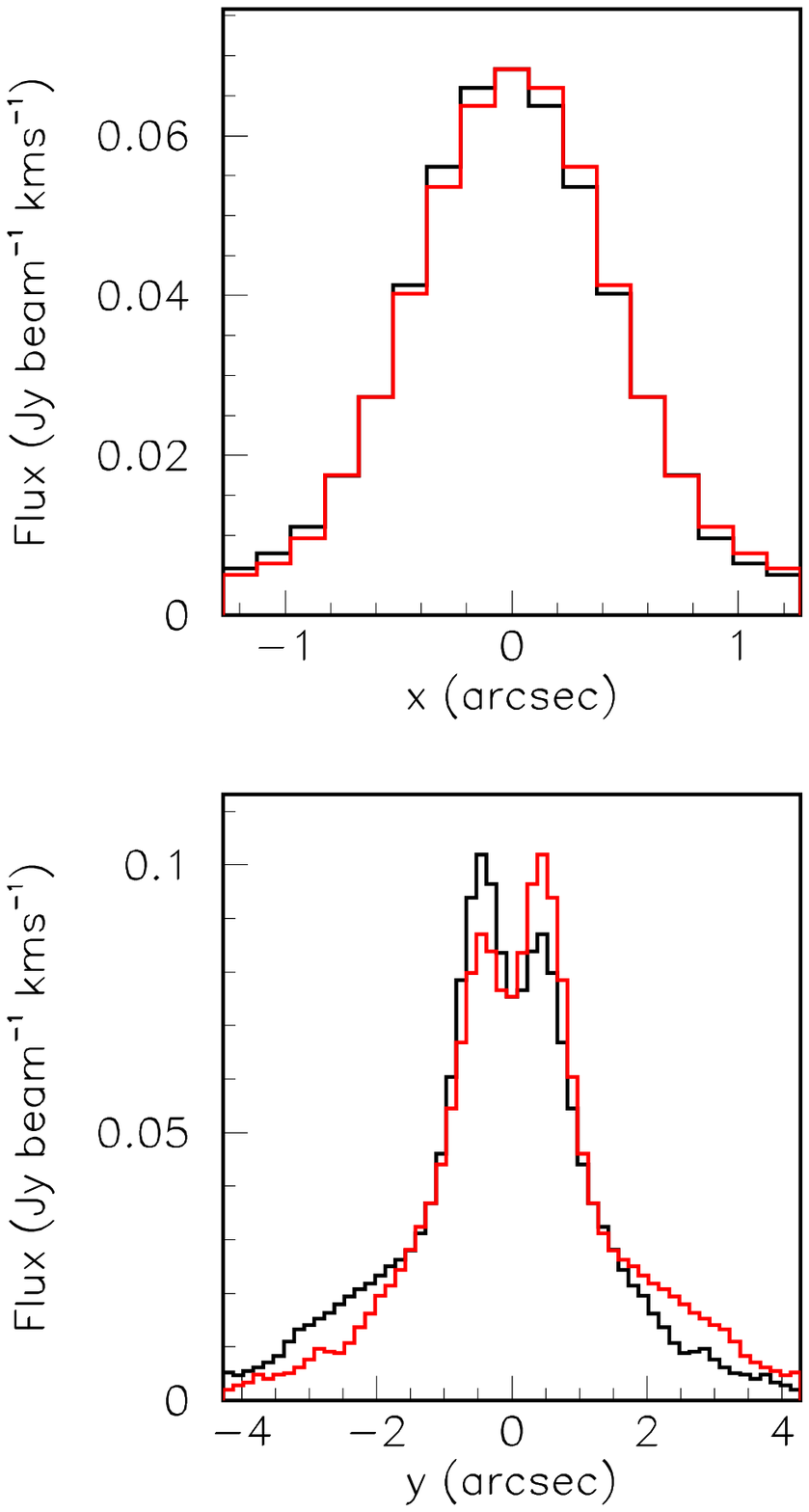}
\caption{Distributions of the flux density \mbox{(Jy beam$^{-1}$ \kms)} integrated over Doppler velocities: on the sky plane (left), projected on $x$ (top right) and on $y$ (bottom right). The disc and its envelope are seen edge-on and project along the $y$ axis while the outflow is centred on the $x$ axis. In the right panels (note the different $x$ and $y$ scales) the red histograms are obtained by symmetry about the origin of coordinates.}
\label{fig1}
\end{figure}

The map of the line flux integrated over Doppler velocities, $F$, is shown in Fig.~\ref{fig1} together with its projections on the $x$ and $y$ axes. The projected distributions are shown together with their reflections with respect to the origin, demonstrating the good agreement between the centres of the continuum and line maps.
A small north-south asymmetry is visible on the $y$ distribution, the southern flux exceeding the northern flux by some 10\% below 1 arcsec and by an average of $\sim$20\% between 1 and 4 arcsec. However, this asymmetry is real and not the result of an offset centre. 

\begin{figure}
\centering
\includegraphics[height=4.2cm,trim=0.5cm 0.3cm 0.5cm 0.3cm,clip]{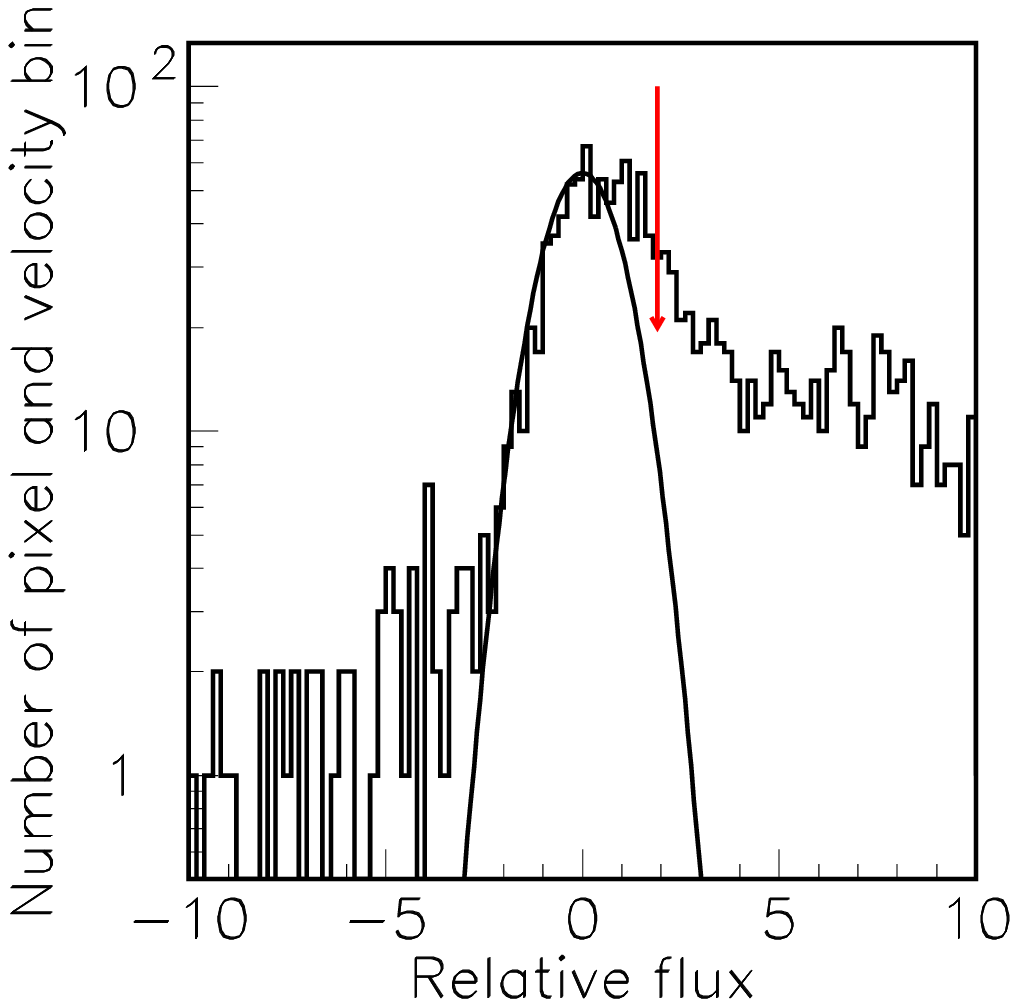}
\includegraphics[height=4.2cm,trim=0.5cm 0.3cm 0.5cm 0.3cm,clip]{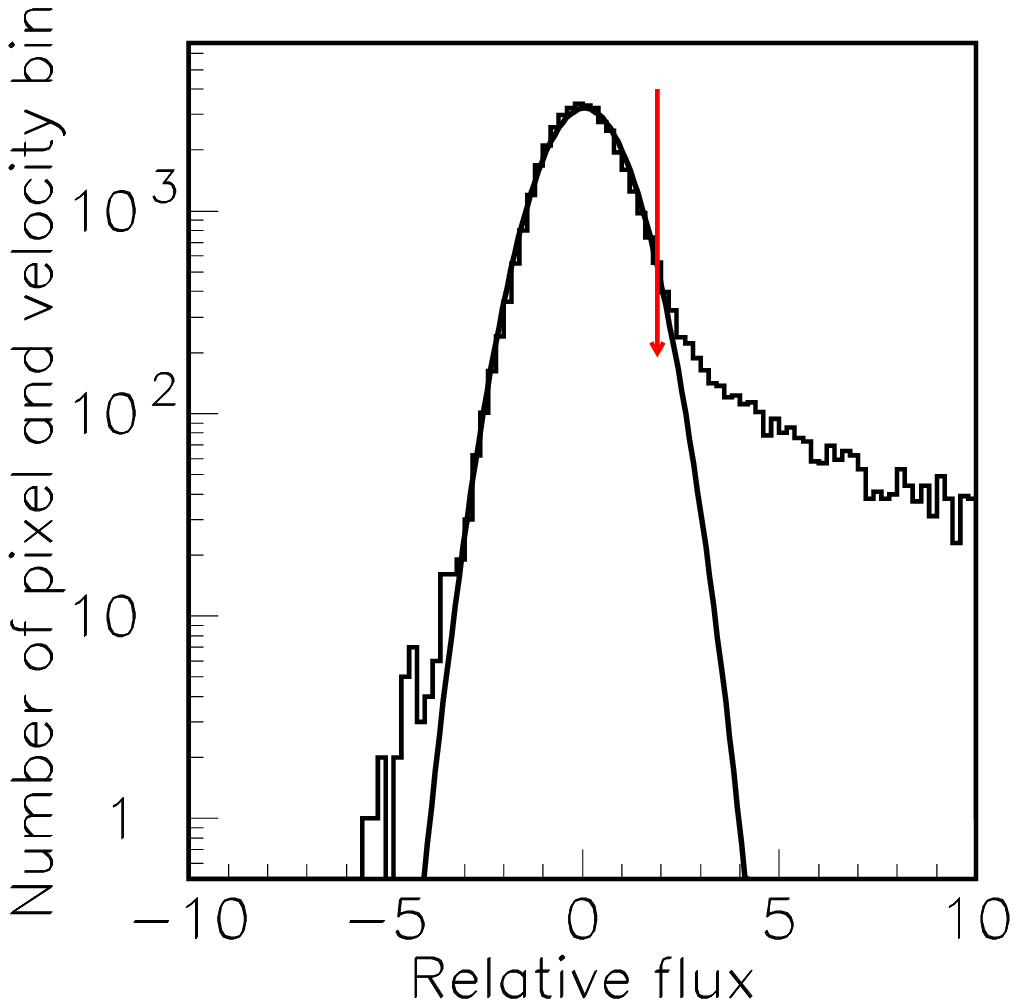}
\caption{Distributions of the flux density divided by the evaluated uncertainty inside (left panel) and outside (right panel)  a rectangle centred at the origin and covering $\pm0.4$ arcsec in $x$ and $\pm0.6$ arcsec in $y$. The curves are Gaussians of unit variance centred at the origin. The red arrows show the position of the 2-$\sigma$ cut.}
\label{fig2}
\end{figure}

We evaluated the position and width of the effective noise, meaning the peak observed at low values in the distribution of the flux density, in each velocity bin separately. In the low velocity bins where the signal is large enough to significantly contribute to nearly all pixels, we extrapolated from the values taken in pixels outside the restricted map used in the analysis. The rms value of the peak varies between 3 and 5 \mbox{mJy beam$^{-1}$} and the offsets between 0 and 5 \mbox{mJy beam$^{-1}$} depending on the intensity of the integrated signal measured in the bin. To each flux density measurement we attach an uncertainty obtained by adding in quadrature the rms of the effective noise and half the value of the offset, the value of the flux density being corrected for the offset. We normally retain for the analysis, unless specified otherwise, flux densities exceeding twice the uncertainty (we speak of a 2-$\sigma$ cut). Figure~\ref{fig2} displays the distributions of the ratio of the flux densities and their associated uncertainties both inside and outside a central rectangle covering 0.8 $\times$ 1.2 arcsec$^2$ in $x \times y$. In the central region, nearly each pixel has a significant content. \\

\section{Overview and main features}

\subsection{Orientation in space}
\begin{figure}
\centering
\includegraphics[width=4.cm,trim=0.3cm 0.cm 0.cm 0.cm,clip]{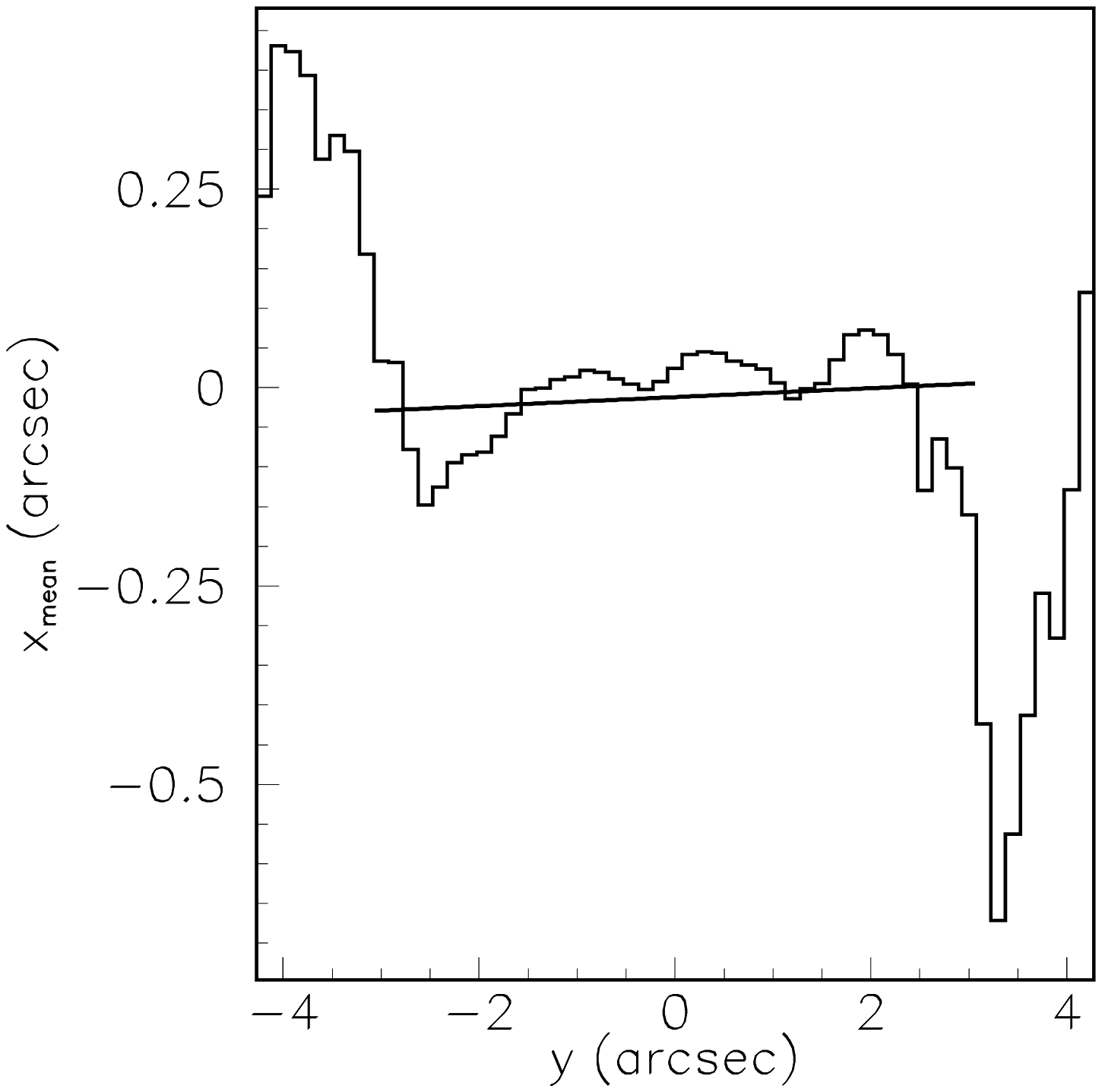}
\includegraphics[width=4.cm,trim=0.3cm 0.cm 0.cm 0.cm,clip]{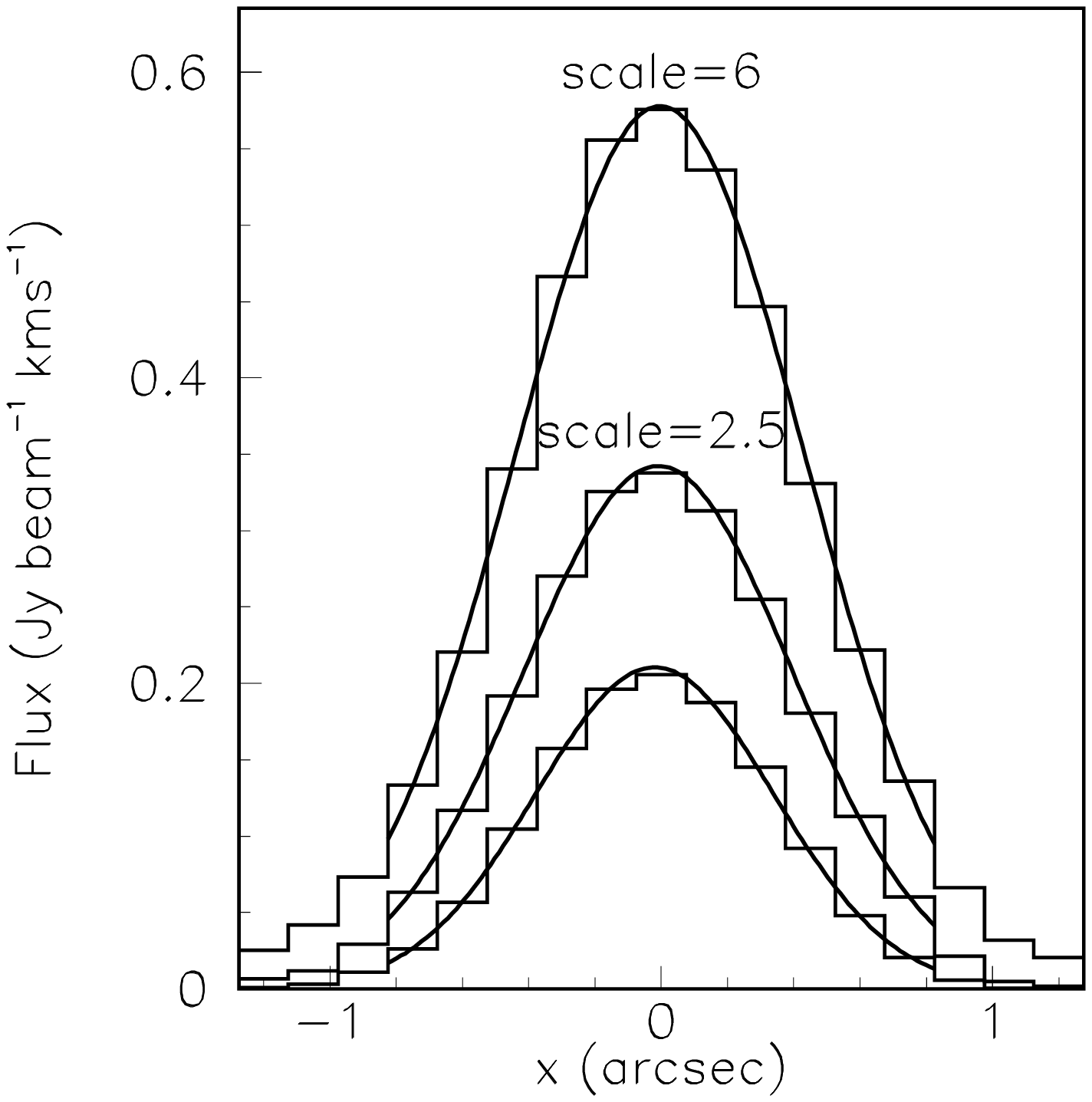}
\caption{Left: distribution of $<x>$, weighted by $F$, as a function of $y$. The line is a linear fit in the interval $|y|<3$ arcsec, of the form $<x>=-0.012$ arcsec$+\,5.7\,10^{-3} y$. Right: distributions of $x$ for $|y|<1$, $2$ and $3$ arcsec from bottom to top. The latter two have been scaled by respective factors of 2.5 and 6 for clarity. The curves are Gaussian fits with FWHM values of $0.85$, $0.96$ and $1.03$ arcsec respectively. Note the different $x$ and $y$ scales.}
\label{fig3}
\end{figure}

As well known from previous studies, the envelope of L1527 is seen edge-on and its axis is oriented east-west, namely along the $x$ axis. In practice, the approximate symmetry plane of the envelope (referred to as the ``disc plane'' for simplicity) may slightly differ from the ($y,z$) plane. A rotation about the line of sight ($z$ axis)  would be immediately visible on the sky map as an angle $\theta_1$ between its approximate symmetry axis and the $y$ axis. A rotation about the $y$ axis by an angle $\theta_2$ is more difficult to reveal and must be obtained from the east-west extension of the measured integrated flux along the $x$ direction, which however combines the effect of $\theta_2$ with those of the beam size and of the intrinsic disk thickness. \citet{Tobin2010} using L' band imaging from the Gemini North telescope, quote a value of 5$^\circ$ for each of $\theta_1$ and $\theta_2$. In the present data, we find that $\theta_1$ does not exceed a degree (Fig.~\ref{fig3} left) when $|y|$ is limited to 3 arcsec but beyond $|y|$ $\sim$3 arcsec, deviations reaching $\sim$0.4 arcsec in $x$ are observed, eastward in the southern hemisphere and westward in the northern hemisphere, corresponding to a counter-clockwise tilt at the 10$^\circ$ scale visible in Fig.~\ref{fig1} (left). The evaluation of the angle $\theta_2$ is obtained from Figure~\ref{fig3} (right), which displays the $x$ distribution of $F$ for $|y|<y_{max}$ for different values of $y_{max}$. In the central region, its FWHM is 0.85 arcsec, compared with 0.47 arcsec for the continuum, meaning $\sim$0.70 arcsec after beam deconvolution. The FWHM of the $y$ distribution is $\sim2.0$ arcsec, meaning $\sim1.9$ arcsec in reality: we may place an upper limit of $\sin^{-1}(0.7/1.9)\sim22^\circ$ on the value of $\theta_2$. In the following we shall usually work in the approximation where both $\theta_1$  and $\theta_2$ cancel.

\begin{figure}
\centering
\includegraphics[height=6.8cm,trim=0.cm 0.cm 0.cm 0.cm,clip]{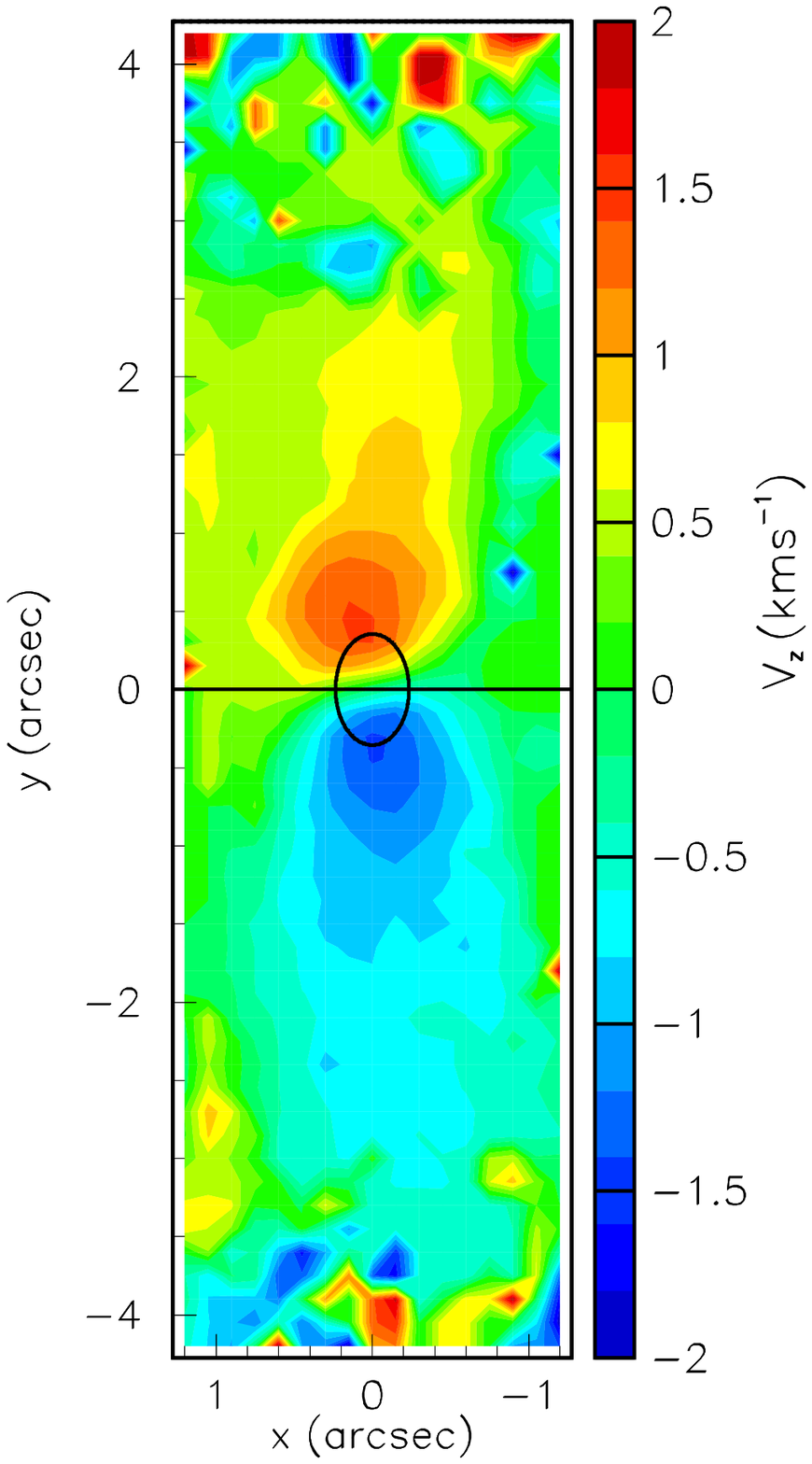}
\includegraphics[height=6.8cm,trim=0.cm 0.cm 0.cm 0.cm,clip]{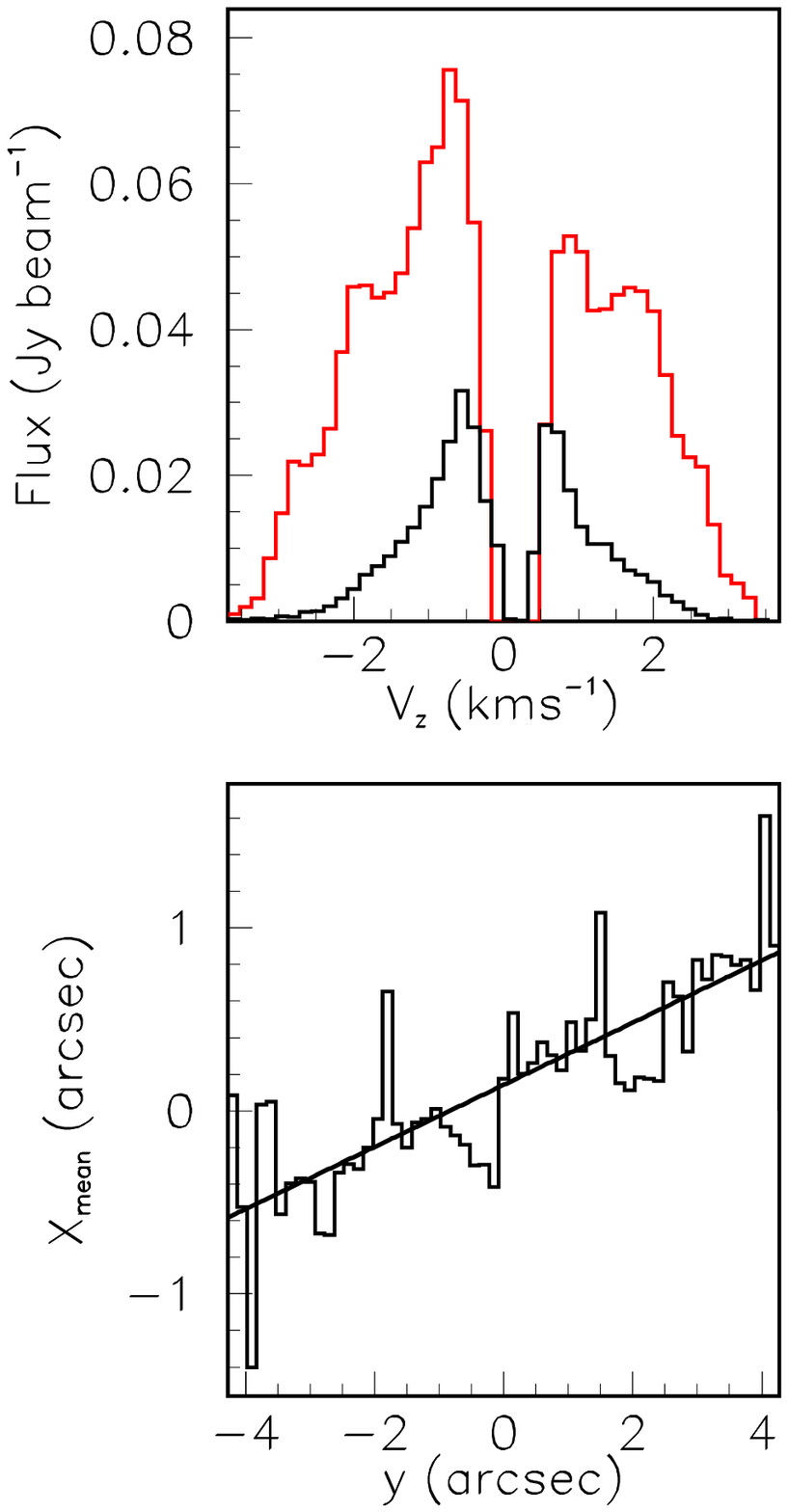}
\caption{Left: sky map of $<V_z>$ (\kms), the ellipse shows the continuum region (FWHM) used in the upper right panel; Upper right: distribution of $F$ on $V_z$ (\kms) inside (red) or outside (black, multiplied by 2 for convenience) the continuum region (FWHM, as sketched in the left panel); Lower right: distribution of $<x>$, weighted by $<V_z>$, as a function of $y$. The line is a linear fit of the form $<x>=0.14$ arcsec$+\,0.17y$.}
\label{fig4}
\end{figure}

\subsection{Rotation}

Figure~\ref{fig4} displays the distribution of the flux density on the Doppler velocity $V_z$ and the sky map of its mean. The dominant feature of the map is the north-south anti-symmetry revealing the well-known rotation of the envelope about the $x$ axis. Closer inspection reveals another significant asymmetry: blue-shifted emission is shifted westward and red-shifted emission eastward (namely having the same inclination with respect to the $y$ axis as the flux excesses noted above for $|y|>3$ arcsec); this is further illustrated in the lower right panel of the figure, displaying the dependence on $y$ of the mean value of $x$ weighted by the mean Doppler velocity in each pixel. A linear fit gives a slope of $-$0.17, corresponding to a tilt of $\tan^{-1}(0.17)=9.6^\circ$ east of north. Namely, while the ``disc planes'' defined by the symmetry of the intensity (Figure~\ref{fig3} right) or of the Doppler velocity (Figure~\ref{fig4} lower right) are both centred at the origin to within one pixel, they differ by about $10^\circ$ in terms of their inclination ($\theta_1$) with respect to the ($y,z$) plane. However, the latter (using velocity) matches the tilt of the former (using intensity) observed at $|y|>3$ arcsec and mentioned in Sub-section 3.1.

\subsection{Absorption}

The Doppler velocity spectrum is split in two by a central dip giving evidence for the major role played by absorption in these observations. Indeed, an optically thin disc of cool gas could not display such a dip. Two explanations have been invoked in earlier publications: absorption of the continuum emission by the foreground gas, extending typically over $\pm0.6$ arcsec in $y$ and slightly red-shifted because of the in-fall contribution to the velocity of the foreground gas; and absorption of the envelope emission by the foreground gas. As the dip is seen both inside and outside the continuum region (Figure~\ref{fig4} upper right), absorption cannot be exclusively blamed on the continuum and an important contribution of extended foreground gas, not only of the envelope, is to be expected \citep{Gueth1997,Guilloteau2016}. While the cool outer envelope, meaning $|y|$>$\sim$0.6 arcsec, is expected to be optically thin at the scale of the present observations, it is seen behind an extended cloud that covers the full map and causes the presence of a dip in the velocity spectrum, again slightly red-shifted because of the average in-falling velocity of the foreground cloud, and extending over the whole $y$ range. 

\begin{figure*}
\centering
\includegraphics[width=.9\textwidth,trim=0.cm 0.cm 0.cm 0.cm,clip]{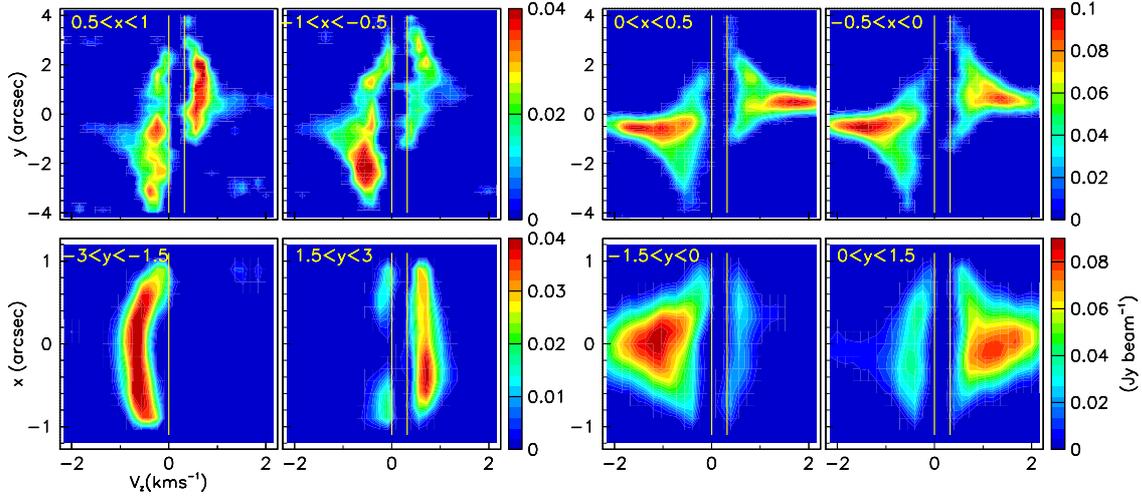}
\caption{P-V diagrams $y$ (up) or $x$ (down) vs $V_z$ in various $x$ (up) or $y$ (down) intervals indicated in each panel. Flux densities are in \mbox{Jy beam$^{-1}$}, angular distances in arcseconds and velocities in \kms. Note the different $x$ and $y$ scales.}
\label{fig5}
\end{figure*}

\begin{figure*}
\centering
\includegraphics[height=7.cm,trim=0.cm 0.cm 0.cm 0.cm,clip]{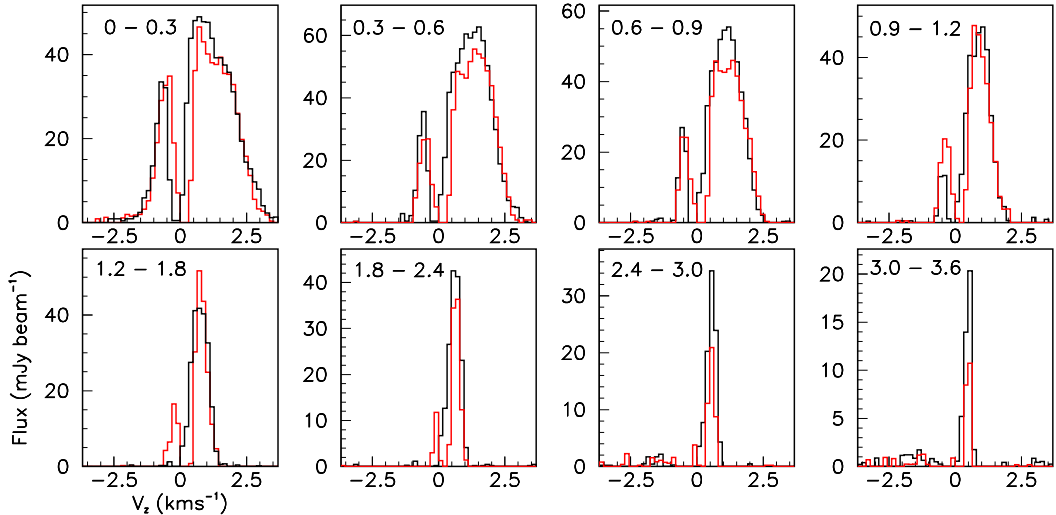}
\caption{Doppler velocity spectra averaged over $|x|<1$ arcsec and intervals of $|y|$ from 0 to $1.2$ arcsec in bins of $0.3$ arcsec and from $1.2$ to $3.6$ arcsec in bins of $0.6$ arcsec. The abscissa is $V_z$ for $y>0$ (red) and $-V_z$ for $y<0$ (black).}
\label{fig6}
\end{figure*}

A closer look at the dependence of $V_z$ on the location of the source is given in Fig.~\ref{fig5}, displaying position-velocity diagrams, with the Doppler velocity as abscissa. The ordinate is one of the sky coordinates and each diagram covers a specified interval of the other coordinate. To first order, they are characteristic of a rotating envelope having its axis along the $x$ axis and displaying symmetries in $x$ (changing $x$ in $-x$ leaves both Doppler velocity and flux densities invariant) and in $y$ (changing $y$ in $-y$ changes the sign of the Doppler velocity and leaves the flux density invariant). But they also reveal the effects of absorption mentioned above: the absence of emission in the two velocity bins above systemic velocity, namely $0<V_z<0.32$ {\kms}, is ubiquitous. To further illustrate this point, we show in Fig.~\ref{fig6} Doppler velocity spectra averaged over $|x|<1$ arcsec and different intervals of $|y|$. Absorption is confined to low velocities, slightly red-shifted. If symmetry in $y$ is obeyed, the velocity spectra measured at two opposite values of $y$ are related by reflection about the origin. This is made clear in the figure by comparing the direct spectrum in the northern hemisphere with the reflected spectrum in the southern hemisphere. If symmetry were obeyed, they would be identical. However, absorption being red-shifted breaks the symmetry: it suppresses the red-shifted part of the direct spectrum and the blue-shifted part of the reflected spectrum, as is clearly observed on the figure.

In what follows we use this feature to select data that are not too strongly affected by absorption. At variance with earlier studies \citep{Ohashi2014} we prefer not to rely on a description of the absorption mechanism, which we would find difficult to model reliably. Adopting such an approach, a detailed study of the spectra displayed in Figure~\ref{fig6} shows that four velocity bins, covering the interval between $-$0.16 and 0.48 {\kms} are essentially lost for analysis. However, if one assumes symmetry under both $x$ and $y$ reflections about the origin, namely invariance of the flux density under transformations of $(x,y,V_z)$ into $(-x,y,V_z)$ and $(x,-y,-V_z)$, the number of lost velocity bins can be reduced to two by retaining the blue-shifted part of the direct spectra and the red-shifted part of the reflected spectra in Fig.~\ref{fig6}, applying small corrections on the neighbouring bins. This feature is being made use of in the following section.

\subsection{Angular dependence of the in-falling flux}

\begin{figure}
\centering
\includegraphics[height=6.cm,trim=0.cm 0.cm 0.cm 0.cm,clip]{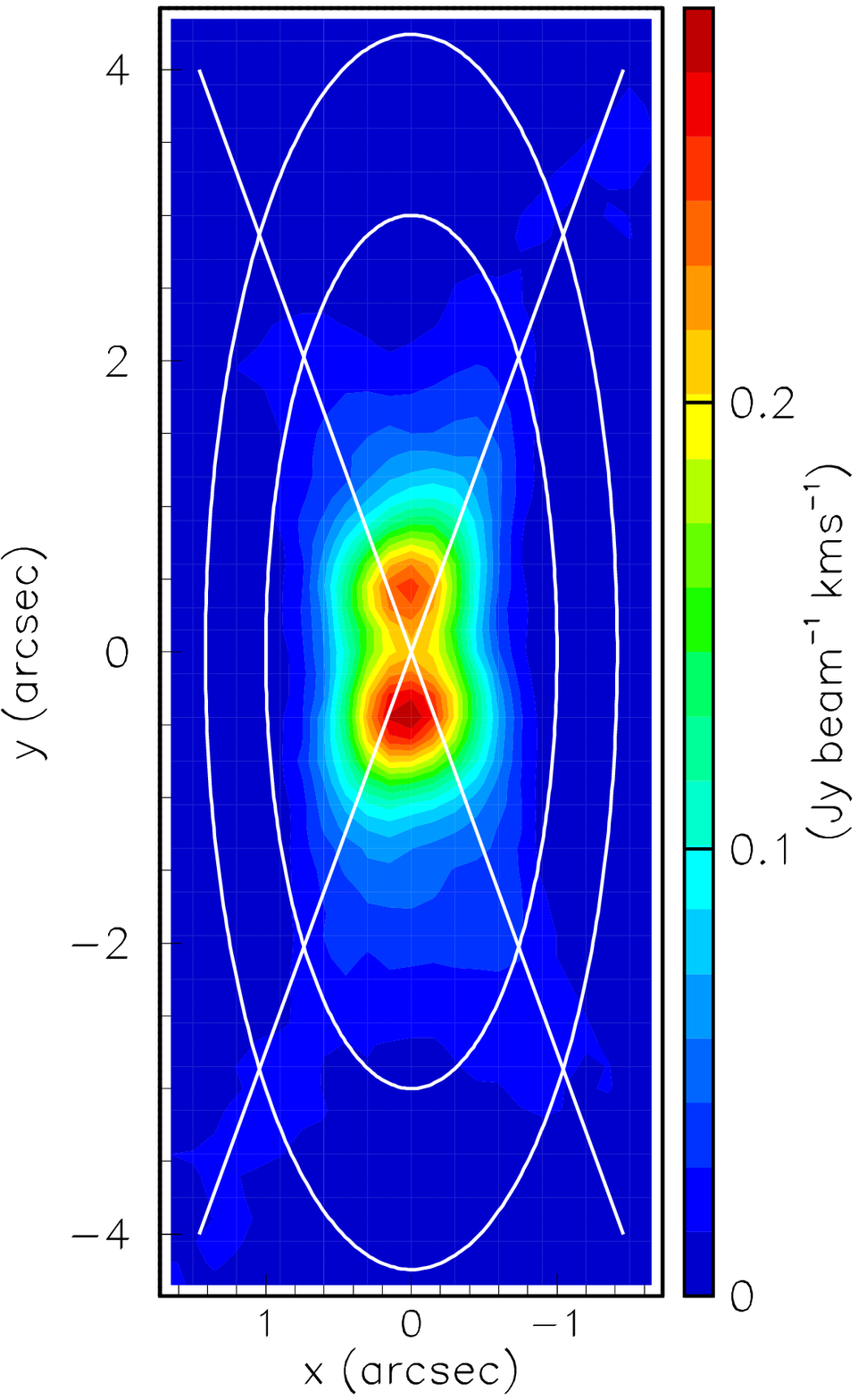}
\includegraphics[height=6.cm,trim=0.cm 0. 0.cm 0.cm,clip]{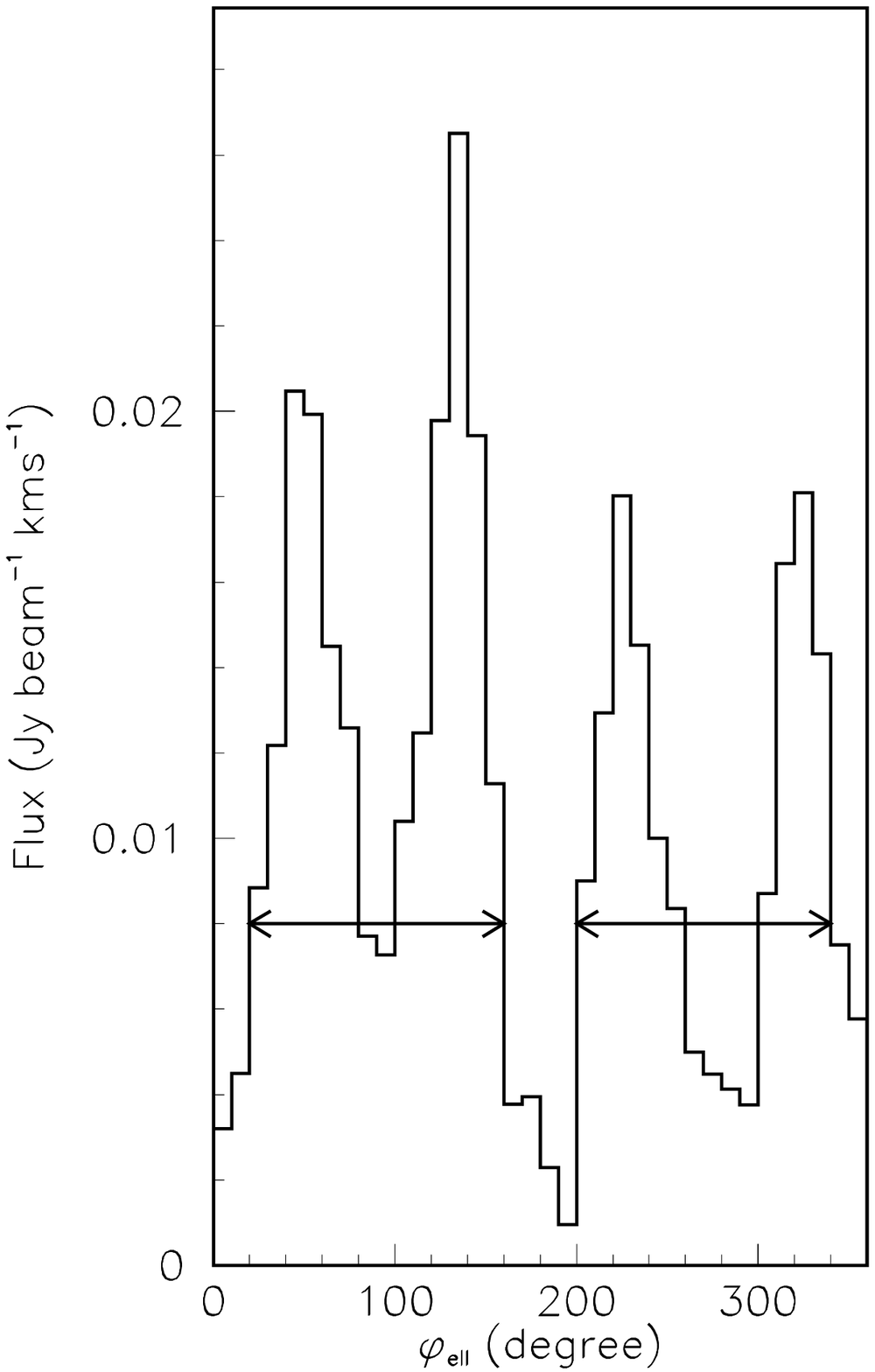}
\caption{Left: Sky map of the integrated flux density showing the elliptical ring used to draw the right panel of the figure. Also shown is the $X$ shape corresponding to $\varphi=\pm70^\circ$ discussed in Sub-section 3.4. Right: Dependence on $\varphi_{ell}$ of the flux density intergrated over $|V_z|<3.68$ \kms and averaged over the elliptical ring displayed in the left panel ($0.8<r_{ell}<1.3$ arcsec).}
\label{fig7}
\end{figure}

Figure~\ref{fig7} (right) displays the dependence of the integrated flux over position angle within an elliptic ring surrounding the star as shown in the left panel. More precisely, we use as abscissa $\varphi_{ell}=\tan^{-1}(^{y}/_{3x})+\pi$ which is simply related to the position angle $\varphi=\tan^{-1}(y/x)+\pi$. The right panel gives direct evidence for the $X$ shape of the flared envelope at large distances visible in the left panel. The in-falling flux is enhanced within $\pm42^\circ$ in $\varphi_{ell}$ from the $y$ axis, meaning within $\pm70^\circ$ in $\varphi$. This direct evidence for an enhancement of the in-falling flux at intermediate values of $\varphi$, or equivalently of a depression of the in-falling flux in the neighbourhood of the disc plane, will be confirmed and discussed in the following sections. 

\section{A simple model of the gas envelope}

\subsection{General description}

We now use the information obtained in the preceding section to produce a simple model of the gas envelope. We assume perfect symmetry in both $x$ and $y$ and rotation invariance about the $x$ axis. The preceding section has shown that these are reasonable approximations and has given quantitative evaluations of the amount by which the data deviate from such an ideal situation. Moreover, we  do not attempt fitting the density and temperature distributions separately; instead we absorb them in a single quantity, the effective emissivity $\rho(x,y,z)$, defined such that 
\begin{equation}
	F(x,y)=\int\rho(x,y,z)dz=\int f(x,y,V_z)dV_z
\end{equation}
where $f(x,y,V_z)$ is the measured flux density and where the $z$ axis is parallel to the line of sight while the $x$ and $y$ axes point east and north respectively. The justification for considering the effective emissivity rather than the temperature and the actual density is twofold: first, having data on a single CO line, we have no simple access to the temperature and second, the important absorption near systemic velocity, which cuts away from analysis a significant fraction of the data, would require a detailed knowledge of the morphology and kinematics of the cloud in which the source is embedded as well as a detailed description of the continuum source, both of which are out of reach. As the C$^{18}$O component is expected to be optically thin, it remains reasonable to assume axisymmetry of the effective emissivity.

The symmetries of the model imply that we may symmetrize the data, averaging flux densities $f(x,y,V_z)$ and $f(-x,y,V_z)$ as well as $f(x,y,V_z)$ and $f(x,-y,-V_z)$. The latter average implies accounting for the slight red-shift of the dip as explained in the preceding section. In practice, two velocity bins are excluded from the analysis, \mbox{$0<V_z<$ 0.32 \kms}, and the contents of the neighbour bins are corrected, the uncertainty attached to each measurement being increased accordingly. The flux densities are evaluated by integration along the line of sight up to 7 arcsec (meaning $\sim$1000 au) in $z$ and distributed among the pixels using a Gaussian beam profile in both $x$ and $y$. The $\chi^2$ is evaluated using as uncertainties the sum in quadrature of the above mentioned uncertainties and 10\% of the measured flux density. We have checked that different reasonable choices essentially lead to the same results as presented below. 

We have paid much attention at introducing as small as possible a number of parameters to describe the morphology and kinematics of the gas volume. To allow the reader who is not interested in details to jump directly to the end of the section (Sub-section 4.3) we describe them briefly in the present paragraph. Guided by elementary hydrodynamics considerations, implying a smooth transition from flared disc to centrally symmetric in-fall, we describe the flux as spiralling on paraboloids having as axis the symmetry axis. A parameter $x_0$ is used to describe their shape and a scaling parameter $\lambda_0$ to label them (increasing for paraboloids becoming less open and more distant from the origin). The total velocity is tangent to the relevant paraboloid and is the sum of a rotation component, normal to the axis, and an in-fall component, contained in the meridian plane ($x,\zeta$) with $\zeta$ pointing along a radius of the disc. The scale of the former is given by a single parameter $V_{rot0}$ and its radial dependence by a power law index going from $-$1 at infinity to $-\frac{1}{2}$ at short distances, the transition being defined by a single parameter, $r_{rot}$. The scale of the latter is given by a single parameter $V_{fall0}$ and its radial dependence by a power law index going to $-m_{fall}$ at large distances and becoming larger at smaller distances as governed by a second parameter $n_{fall}$. In practice, we find that both $m_{fall}$ and $n_{fall}$ are very poorly defined and that the data are satisfied with a nearly constant in-fall velocity. An important result of our analysis is the evidence for a significant depression of the effective emissivity near the disc plane as previously mentioned in Section 3.4. In order to parameterise its shape, we need two parameters, $\lambda_1$ measuring its extension around the disc plane and $\epsilon$ measuring its scale. The stationary isothermal regime that can be described by the above parameters is used here as a reference, but is obviously improper to describe reality. An additional parameter $q$ is introduced to measure the deviation of the flux from such an ideal regime. A last parameter $\lambda_{rot}$ describes the observed dependence of the rotation velocity when moving away from the ($y,z$) plane (Figure~\ref{fig5} lower left).

In total, the model uses eleven parameters in addition to the scale factor $\rho_0$ readily obtained by normalisation of the integrated flux predicted by the model to the measured flux. Of these eleven parameters, two, $m_{fall}$ and $n_{fall}$ turnout to be so poorly constrained that they could as well be omitted and one, $\lambda_{rot}$, is trivially obtained from the data. We are therefore dealing effectively with a 8-parameter model to describe the morphology and kinematics of the envelope. 

\subsection{Parameterisation}

\begin{figure*}
\centering
\includegraphics[height=6.cm,trim=0.cm 0.5cm 0.cm 0.5cm,clip]{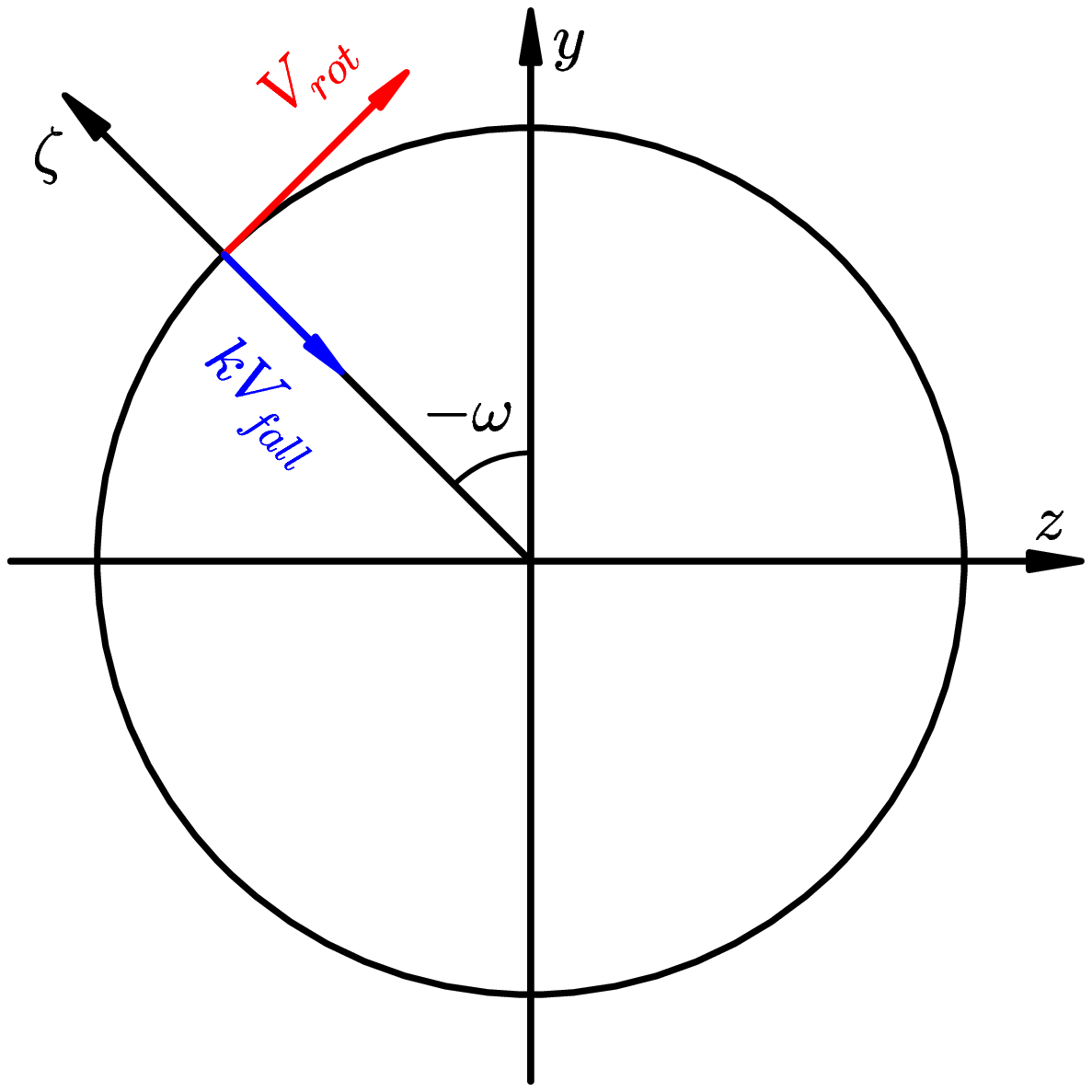}
\includegraphics[height=6.cm,trim=1.cm 0.8cm 0.cm 0.5cm,clip]{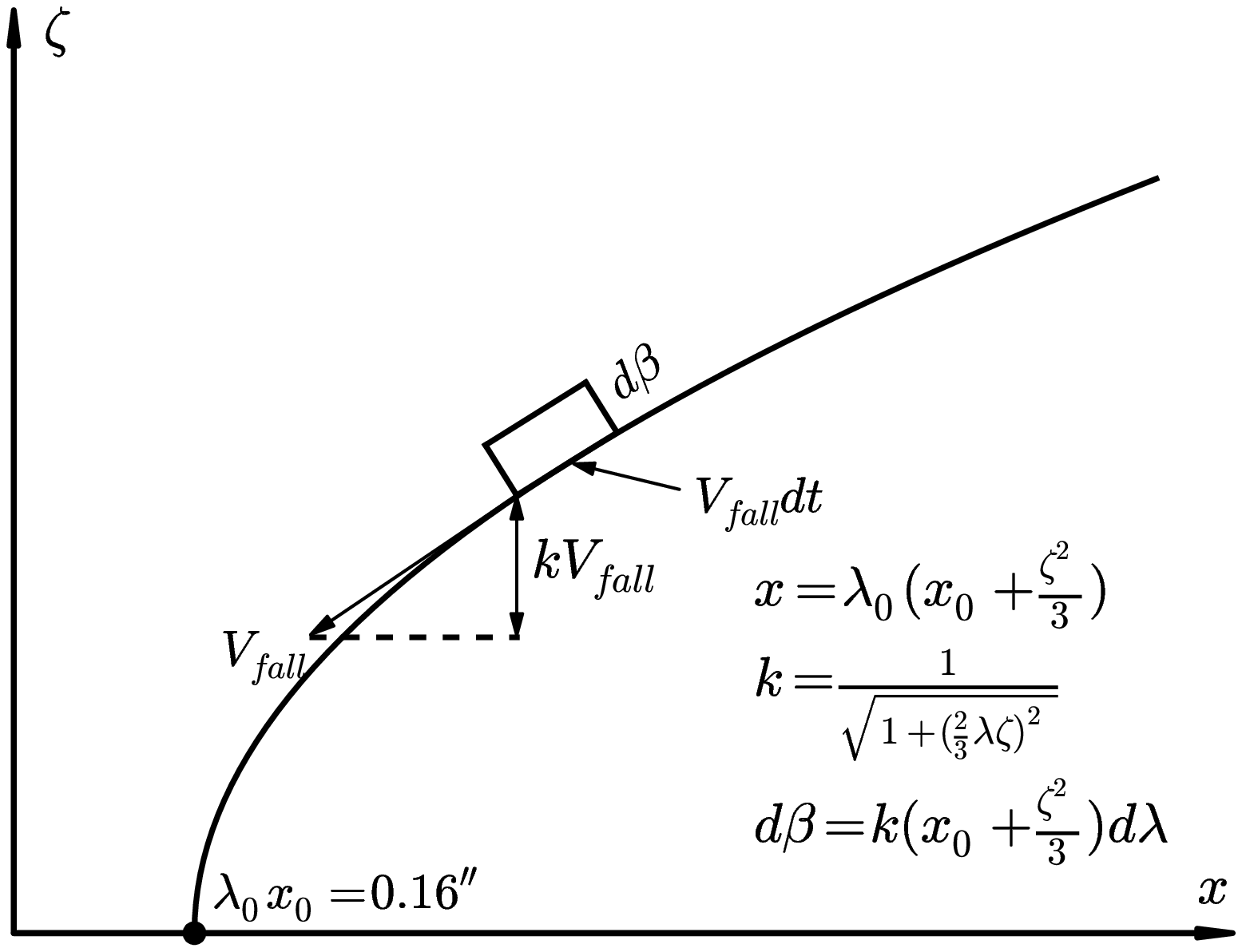}
\caption{Geometry of the model. Left: In the $(y,z)$ plane. The properties of the envelope are independent of $\omega$, namely invariant by rotation about the $x$ axis. The $\zeta$ axis is along any radius of the disc. The rotation velocity $V_{rot}$ is parallel to the $(z,y)$ plane and the projection of the in-fall velocity on this plane is $kV_{fall}$ pointing to the origin. Right: On the $(x,\zeta)$ plane, showing the parabolic trace of the paraboloid having $\lambda=\lambda_0$. The rotation velocity is perpendicular to the plane and the in-fall velocity is tangent to the parabola with component $kV_{fall}$ along $\zeta$. The insert sketches the arithmetic that applies to the case of stationary flow.}
\label{fig8}
\end{figure*}

We model (Fig.~\ref{fig8}) the trajectories of the gas molecules as spiralling on paraboloids of equation \mbox{$x=\lambda(x_0+\zeta^2/(3$ arcsec))} with the cylindrical radius \mbox{$\zeta=\sqrt{y^2+z^2}$} (note that the choice of 3 arcsec in the denominator is arbitrary).

The velocity is parameterised as the sum of a component $V_{rot}$ perpendicular to the ($\zeta$,$x$) plane and a velocity $V_{fall}$ tangent to the parabola in the ($\zeta$,$x$) plane. If $\omega=\tan^{-1}(z/y)$ is the angle between the ($\zeta$,$x$) plane and the plane of the sky, the component of $V_{rot}$ on $z$ is $V_{rot}\cos\omega$. The unit vector tangent to the parabola has components $\sfrac{2}{3}k\lambda\zeta$ on $x$ and $k$ on $\zeta$ with $k=[1+(\sfrac{2}{3}\lambda\zeta)^2]^{-\sfrac{1}{2}}$. The component of $V_{fall}$ on $z$ is $-kV_{fall}\sin\omega$. Hence 
\begin{equation}
	V_z=V_{rot}\cos\omega-kV_{fall}\sin\omega.
\end{equation}
Our hypothesis of rotational invariance implies that the morphology and kinematics of the envelope are independent of $\omega$. We parameterise both $V_{rot}$ and $V_{fall}$ as functions of the distance to the origin, \mbox{$r=(\zeta^2+x^2)^{\sfrac{1}{2}}$}. The rotation velocity is known to evolve from an inverse distance law at large distances from the protostar to Keplerian at short distances \citep{Tobin2012, Ohashi2014}, which we translate as a power law in $r$ of index $n_{rot}=\sfrac{1}{2}\,\exp(-r/r_{rot})-1$, giving $n_{rot}=-\sfrac{1}{2}$ at $r=0$ and $n_{rot}=-1$ at $r=\infty$ with $r_{rot}$ measuring the distance from the protostar at which the transition takes place.  Moreover, we allow for a small decrease of the rotation velocity when moving away from the $(y,z)$ plane ($x$=0) as seen in the lower left panel of Fig.~\ref{fig5} and we write $V_{rot}=V_{rot0}\,\exp(-\sfrac{1}{2}\lambda^2/\lambda^2_{rot}) r^{n_{rot}}$.

The velocity $V_{fall}$ must cancel when approaching the disc and behave as a power law at large distances, which we write as 
\begin{equation}
	V_{fall}=V_{fall0}\,r^{n_{fall}}/(1+r^{n_{fall}+m_{fall}}) 
\end{equation}
reaching its maximum at $r_{fall}=(n_{fall}/m_{fall})^{1/(n_{fall}+m_{fall})}$ and behaving as $r^{-m_{fall}}$ at large distances. Note that at variance with earlier models \citep{Ohashi2014, Sakai2014a, Sakai2014b} our evaluation of the in-fall velocity rests exclusively on the kinematics of the envelope, without any constraint from the absorption.

We parameterise the density $d$ as a multiple of the density $d_{stat}$ associated with a constant accretion rate $dM/dt=\int A(\lambda)d\lambda$ in a stationary regime. At a distance $\zeta$ from the $x$ axis, the mass of gas flowing in a time $dt$ across $\zeta$ between paraboloids $\lambda$ and $\lambda+d\lambda$ is 
\begin{align}
A(\lambda)d\lambda dt=2\pi\zeta d_{stat} V_{fall} dt d\lambda(x_0+\zeta^2/(3\,\textrm{arcsec}))k \\
\mbox{Hence } d_{stat}=A(\lambda)/[2\pi\zeta V_{fall}(x_0+\zeta^2/(3\,\textrm{arcsec}))k].
\end{align}
To parameterise $A(\lambda)$, we remark that it must decrease rapidly when $\lambda$ increases in order to leave room for the broad outflow about the $x$ axis. However, a Gaussian form $A(\lambda)=A_0 exp[-\sfrac{1}{2}(\lambda/\lambda_0)^2]$ cannot reproduce the depression observed at small values of $\lambda$ near the plane of the disc (Fig.~\ref{fig7} right). We write therefore
\begin{equation}
A(\lambda)=A_0\{\exp[-\sfrac{1}{2}(\lambda/\lambda_0)^2]-\epsilon\,\exp[-\sfrac{1}{2}(\lambda/\lambda_1)^2]\}
\end{equation}

At constant temperature and in a non-turbulent optically thin regime in local thermal equilibrium, the effective emissivity $\rho$ is simply proportional to $d_{stat}$. In order to account for deviations from such an ideal situation, we simply include a factor $r^q$ in the denominator of  $d_{stat}$ with $q$ measuring the importance of such deviations, namely
\begin{equation}
\rho(\lambda)=\frac{\rho_0\{\exp[-\sfrac{1}{2}(\lambda/\lambda_0)^2]-\epsilon\,\exp[-\sfrac{1}{2}(\lambda/\lambda_1)^2]\}}{2\pi\zeta r^q V_{fall}(x_0+\zeta^2/3)k}
\end{equation}

\subsection{Result}

\begin{figure*}
\centering
\includegraphics[height=6.5cm,trim=0.cm 0.cm 0.cm 0.cm,clip]{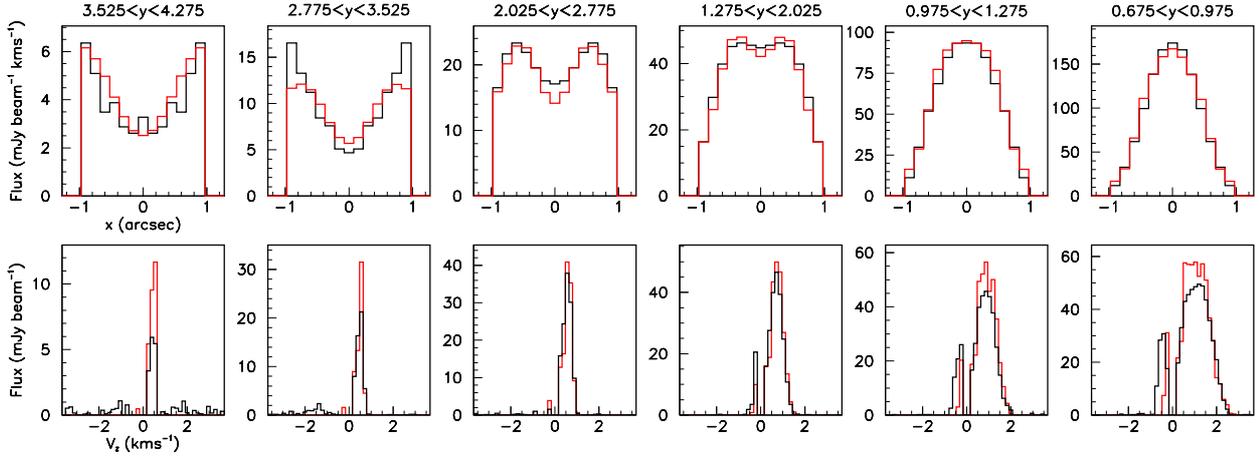}
\caption{ Measured and symmetrized (black) and model best fit (red) distributions of the integrated flux are shown as a function of $x$ in the upper row and as a function of $V_z$ in the lower row for different intervals of $|y|$ (as labelled above the upper row). In the lower row data having $y<0$ are plotted as a function of $-V_z$. }
\label{fig9}
\end{figure*}

\begin{figure*}
\centering
\includegraphics[height=4.5cm,trim=0.cm 0.cm 0.cm 0.cm,clip]{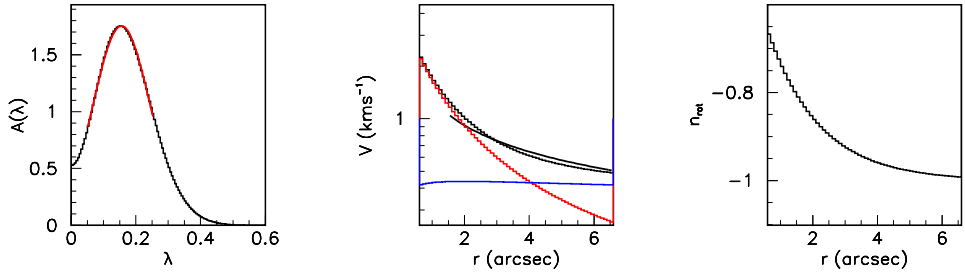}
\caption{Best fit values of the model parameters. Left: Dependence on $\lambda$ of $A(\lambda)$. The red curve is a Gaussian fit for $0.05<\lambda<0.25$ giving a mean of 0.156 and an rms of 0.088. Middle: Dependence on $r$ of $V_{rot}$ (red, for $\lambda$=0), $V_{fall}$ (blue) and \mbox{$V_{tot}=\sqrt{V_{rot}^2+V_{fall}^2}$ (black)}. The curve is the result of a fit to a $r^{-\sfrac{1}{2}}$ law for $r>1.5$ arcsec giving $(2GM)^{\sfrac{1}{2}} =1.3$ \kms arcsec$^{\sfrac{1}{2}}$ (see text). Right: Dependence on $r$ of $n_{rot}$.}
\label{fig10}
\end{figure*}

\begin{figure*}
\centering
\includegraphics[height=5.cm,trim=0.cm 0.cm 0.cm 0.cm,clip]{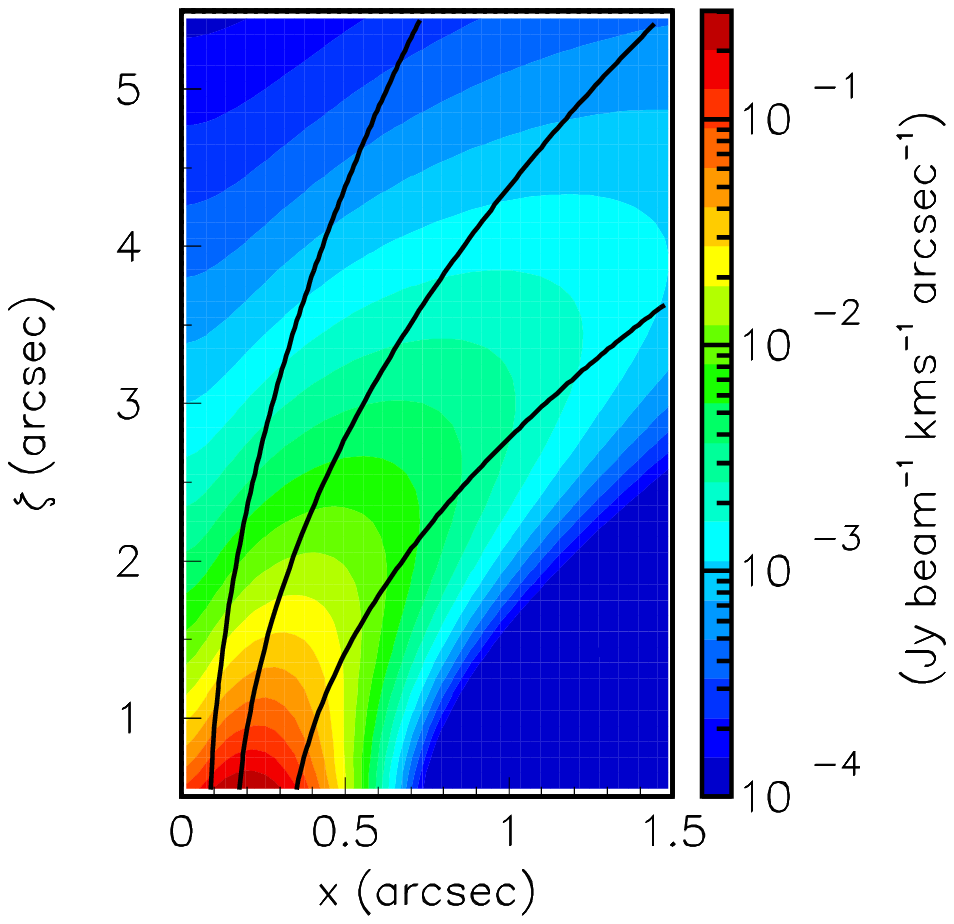}
\includegraphics[height=5.cm,trim=0.cm 0.cm 0.cm 0.cm,clip]{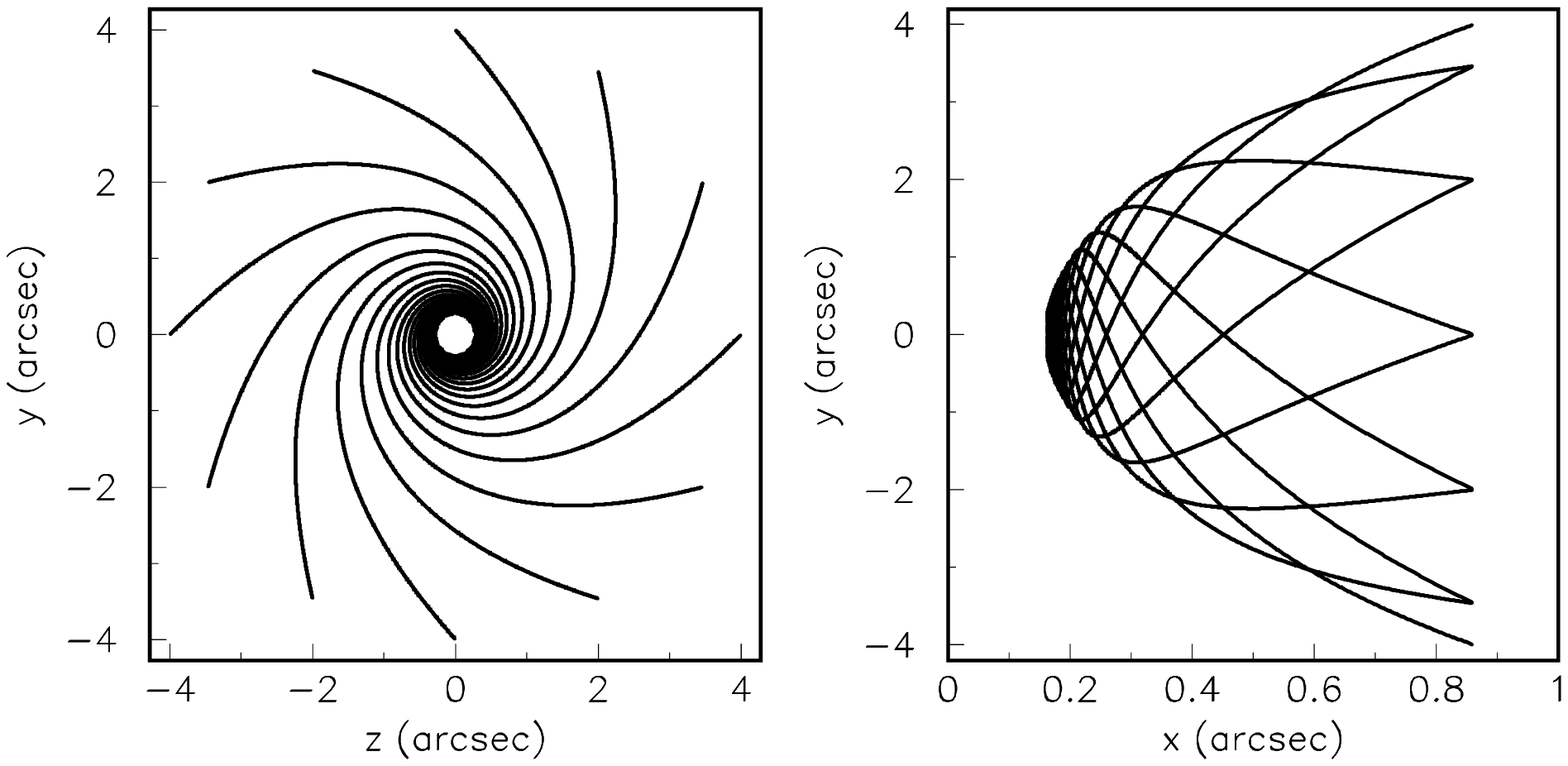}
\caption{Left: map of the effective emissivity in the ($\zeta$,$x$) plane. Note the different $x$ and $\zeta$ scales. The three parabolas are for $\lambda=\sfrac{1}{2}\lambda_0$, $\lambda_0$ and $2\lambda_0$. Middle and right: projection on the ($y$,$z$) and ($y$,$x$) planes respectively of model trajectories having $\lambda=\lambda_0$ and starting at $\zeta=4$ arcsec with values of $\omega$ at $30^\circ$ intervals.}
\label{fig11}
\end{figure*}

The result of the best fit is summarized in Table~\ref{tab_bestfit} and illustrated in Figures~\ref{fig9} to \ref{fig11}. The quantities $\Delta^\pm$ listed in the table are amounts by which the parameter has to increase ($\Delta^+$) or to decrease ($\Delta^-$) for $\chi^2$ to increase by 5\%, other parameters being kept constant. They should not be rigorously understood as real uncertainties but as giving an evaluation of the sensitivity of the quality of the fit to each parameter separately. Yet, their definition has been chosen in such a way as to give a sensible estimate of the actual uncertainties.

\begin{table*}
\centering  
\caption{Best fit values of the model parameters. Distances are measured in arcseconds and velocities in \kms.}    
\label{tab_bestfit}        
\begin{tabular}{ccccc}
\hline 
 & Parameter & Best fit value & $\Delta^-$ & $\Delta^+$ \\
\hline 
$x_0$ & Offsets on $x$ axis & 1.24 & 0.20 & 0.31 \\
$\lambda_0$ & Paraboloid label & 0.131 & 0.015 & 0.016 \\
$\epsilon$ & Depth of median depression & 0.940 & 0.060 & 0.043 \\
$\lambda_1$& Width of median depression & 0.102 & 0.035 & 0.017 \\
$q$ & Non-stationarity & 0.93 & 0.25 & 0.24 \\
$V_{rot0}$ & Rotation velocity & 1.66 & 0.06 & 0.09 \\
$r_{rot}$ & Transition $1/r$ to Keplerian & 1.60 & 0.93 & 0.70 \\
$\lambda_{rot}$ & Slowing down rotation vs $x$ & 0.28 & 0.03 & 0.09 \\
$V_{fall0}$ & In-fall velocity & 0.86 & 0.17 & 0.23 \\
$n_{fall}$ & Slowing down in-fall vs $r$ & 0.32 & 0.32 & 0.48 \\
$m_{fall}$ & Power index in-fall & 0.22 & 0.22 & 0.16 \\
\hline
\end{tabular} 
\end{table*}

Figure \ref{fig9} illustrates the quality of the fit, both the integrated flux and the Doppler velocity distributions are rather well reproduced in the domain covered by the model ($|x|<0.975$ arcsec; $0.675<|y|<4.275$ arcsec; |$V_z-$0.16\,\kms|>0.16\,\kms). The figure makes full use of the symmetries of the model by displaying symmetrized data. Figure~\ref{fig10} displays the dependences on $r$ and $\lambda$ of the main parameters of the model. Figure~\ref{fig11} shows the map of the space-reconstructed effective emissivity in the \mbox{($x$,$\zeta$)} plane as given by the best fit and illustrates typical spiral trajectories. 

\section{Discussion}

\begin{figure}
\centering
\includegraphics[height=6.cm,trim=0.cm 0.cm 0.cm 0.cm,clip]{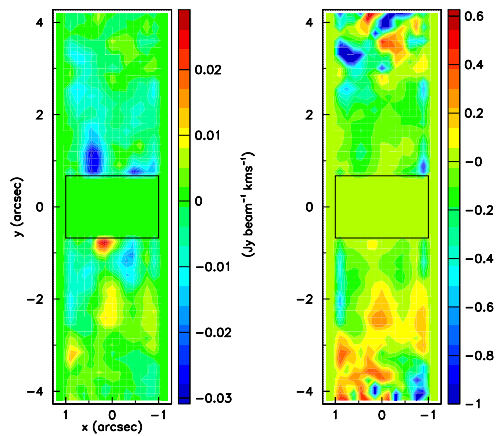}
\caption{Deviations between the values taken by the integrated flux over the sky map in the data and the model. The left panel displays absolute deviations, $F_{data}-F_{model}$ and the right panel displays asymmetries, $(F_{data}-F_{model})/(F_{data}+F_{model})$. The rectangles show the central region excluded from the fit.}
\label{fig12}
\end{figure}

The results obtained in the preceding section are qualitatively consistent with earlier descriptions of the protostar envelope but contribute significant new quantitative information on its morphology and kinematics. Figure~\ref{fig12} displays maps of the deviation between the model and the data, both in absolute value (right panel) and as an asymmetry (left panel). The former are dominated by pixels close to the limit of the central region excluded from the fit while the latter do not exceed $\sim$20\% for $|y|<$3 arcsec. Beyond this value, the individual pixel contents are close to noise level. In general, the maps illustrate the good quality of the fit given the very simple form of the model.

\subsection{Rotation}
Over most of the explored range the kinematics is dominated by rotation, with an accurately measured velocity. The choice of having velocities tangent to paraboloids is very well accommodated by the data, attempts to use different shapes did not bring any significant improvement to the quality of the fit. However, we cannot claim evidence for this particular choice of trajectories being unique. The limitation of the analysis to values of $|y|$ exceeding $\sim$100 au prevents a reliable exploration of the central region where the Keplerian disc is located. While the need for $n_{rot}$ to increase toward small $r$ values is clearly required by the data, the value of $r_{rot}$, $\sim$1.6$\pm$0.9 arcsec, an approximate measure of the distance from the star where the transition to a Keplerian regime occurs, is poorly constrained: the existence of a Keplerian disc having a diameter at the arcsecond scale is consistent with the present observations, which, however, do not provide a more accurate evaluation of its dimensions. Extending the analysis to lower values of $|y|$ would require a reliable modelling of the emission and absorption properties of the protostar, the disc and the molecular cloud in which they are embedded.
Evidence for an additional small decrease of the rotation velocity when moving away from the disc plane, already apparent in the raw data (Fig.~\ref{fig5} lower left), is confirmed by the fit with a $\sigma$ of $\lambda_{rot}$=0.28, meaning 1.2 arcsec (2.8 arcsec FWHM) at a distance $\zeta$=3 arcsec from the disc axis.

\subsection{In-fall velocity}
The sensitivity of the present observations is sufficient to give evidence for in-fall, but insufficient to measure its variation precisely over the explored domain: the best fit values of $V_{fall}$ are confined between 0.44 and 0.41 \kms when $r$  spans between 1 and 5 arcsec. Namely, we may retain a mean in-fall velocity of 0.43$\pm$0.10 \kms over the explored range with no evidence for a significant $r$-dependence. \citet{Ohashi2014} did not adjust the in-fall velocity to best fit their data but adopted a different approach: having remarked that setting the in-fall velocity to its free fall value gives a bad fit to the data, they reduced it by a factor of 2 for $r>$1.8 arcsec (to better match their CO data) and by a factor of 5 for $r<$0.9 arcsec (to better match their SO data) and were satisfied with the result. In the range $r>$4 arcsec, where the total velocity is not too strongly dominated by the rotation velocity, therefore where the in-fall velocity can be best measured, the value assumed by \citet{Ohashi2014} agrees well with that obtained from the present analysis. Good agreement is also obtained at small values of $r$ where both analyses require the in-fall velocity to tend to zero. However, the sudden jump postulated by \citet{Ohashi2014} in the 0.9$<r<$1.8 arcsec interval, reaching high values of ~1.5 \kms, is not required by the present analysis. \citet{Ohashi2014} need such a jump to reproduce the velocity profile measured at the central star position (0.75$\times$0.75 arcsec$^2$). However, the fit that they obtain is poor (their Figure 1c). Moreover the central star region is strongly influenced by absorption and has been excluded from the present analysis as being too difficult to describe reliably. 
		
\subsection{Central mass}
The space velocity, $V_{tot}=\sqrt{V_{rot}^2+V_{fall}^2}$, increases when $r$ decreases as shown in Fig.~\ref{fig10} (middle). Writing that the total energy is conserved and cancels at infinity provides an estimate of the mass $M$ contained in the central region. The measured velocity $V_{tot}$=1.3 \kms at $r$=1.5 arcsec being then equal to $\sqrt{2GM/r}$ (where $G$ is Newton's gravitation constant) we measure $M\sim$0.2 solar masses. This measurement agrees well with the evaluation made by \citet{Tobin2012} using $^{13}$CO (2-1) CARMA observations. They find $M$=0.19$\pm$0.04 solar masses from the rotation curve measured at distances from the star not exceeding 100 au (0.7 arcsec), namely below the range explored in the present analysis. The quoted error reflects only the quality of the fit but excludes systematic uncertainties attached to the reliability of the method.  
However, when fitting $V_{tot}$ to a power law in $r$ over the whole observed range, we find a much better fit for a power index of $-$0.7 instead of $-$0.5 as required by the energy conservation argument quoted above. This was also noted by \citet{Ohashi2014}. It is to be seen in conjunction with the important deviation from zero measured for $q$, 0.93$\pm$0.25, illustrating the approximate nature of the assumptions made in constructing and/or interpreting the model (no viscosity, stationarity, no turbulences, thermal equilibrium, optical thinness, etc.). In particular, a non-zero value of $q$ is straightforwardly generated by variations of the temperature with $r$. This suggests, rather than comparing kinetic energy with potential energy, to compare instead their variations with $r$. Writing that the kinetic energy increase is compensated by an equal decrease of the gravitational potential energy at $r$=1.5 arcsec, we find a value of $M$ 0.7/0.5 times larger than previously obtained, namely $\sim$0.3 solar masses. The very approximate nature of these evaluations must be kept in mind and we retain an estimate of 0.23$\pm$0.06 solar masses contained within $r<$1.5 arcsec, where the uncertainty is meant to account for systematic contributions. Possible mechanisms causing the in-fall velocity to be significantly smaller than the free-fall velocity have been reviewed in the earlier literature. They do not need to be repeated here as the present analysis does not provide new arguments that could favour one of these over the others. 

\subsection{Disc plane depression}
The dependence on $\lambda$ of the accretion rate is accurately measured and could as well have been modelled by a Gaussian enhancement in $\lambda$ having a mean value of $\lambda_{max}$=0.156 and a $\sigma$ of 0.088 (Fig.~\ref{fig10} left). There is clear evidence for a strong depression in the disc plane as well as over a broad solid angle about the disc axis where outflow is known to take place. We prefer to talk of a depression in the disc plane rather than an $X$-shaped enhancement because simple gravity requires the density to increase toward the disc plane. The geometry of the observed depression makes it very unlikely that it be the result of an artefact of the model or of opacity. It is therefore reasonable to assume that it reveals an actual depression of the CO(2-1) emission. As well-known \citep[for a recent review see][]{Caselli2012} the proximity of the gas temperature to the CO sublimation temperature implies that freeze-out of CO molecules on cold dust grains - or conversely sublimation - is a dominant process in the dynamics that govern the evolution of gas envelopes. Many astrochemical models predict a protoplanetary disc structure similar to what is found here \citep[for a recent example see][]{Fogel2011}. Another example of the importance of the role played by CO freeze-out on (or sublimation from) dust grains is a recent study \citep{Jorgensen2015} suggesting that protostars such as L1527 undergo significant bursts once every 20'000 years causing sublimation of the frozen-out CO molecules. This suggests an interpretation of the observed depression in terms of CO molecules freezing out on cold dust preferably near the median disc plane. Ultimately \citep[see Figure 14 of][]{Caselli2012} this is indeed what is expected and it is not unreasonable to assume that it is starting to take place in the case of L1527.

\subsection{Accretion rate}
The normalisation between the effective emissivity obtained in the model and observed in the data, fixing $\rho_0$, gives an evaluation of the accretion rate of 3.5 10$^{-7}$ \Msold, where we assume a constant temperature of 30 K and a C$^{18}$O to molecular hydrogen ratio of 4.9 10$^{-8}$ \citep{Jorgensen2002}. This number is evaluated at a distance of 1 arcsec from the star. A significantly higher accretion rate of 6.6 10$^{-7}$ \Msold was suggested by \citet{Tobin2012} using luminosity arguments. However, this evaluation is very indirect and the uncertainty attached to it is large. Rather than being a cause of concern, the fact that the two numbers differ by a factor of nearly 2 suggests that the associated methods measure different quantities that are subject to very different systematic uncertainties. In particular, the evaluation obtained from the present analysis depends on the distance from the star, decreasing by a factor $2^{0.93}$ when moving away to a distance of 2 arcsec. Taking this into account, we retain a value of 3.5$\pm$1.0 10$^{-7}$ \Msold at a distance of 1 arcsec (140 au) from the star.

\subsection{Orientation in space}
The envelope thickness at $y$=0, measured by a FWHM of $\sim$0.55 in $\lambda$ (Fig.~\ref{fig10} left) meaning $\sim$0.7 arcsec, is consistent with the value quoted in Sub-section 3.1 from which an upper limit of $\sim22^\circ$ was placed on a possible tilt $\theta_2$ between the disc plane and the $(y,z)$ plane about the $y$ axis. The observed $\sim10^\circ$ tilt $\theta_1$ about the $z$ axis of the map of $<V_z>$ that was mentioned in Sub-section 3.2 and illustrated in Fig.~\ref{fig4} deserves some comments. As remarked then, such a tilt is also seen in the integrated flux for distances from the star exceeding $\sim$3 arcsec, not only in the present data (Figures~\ref{fig1} left and~\ref{fig3} left), but also in infrared observations \citep{Tobin2008}. The difference of inclination of the central part of the envelope, as measured from the integrated flux, and of the outer part, including the outflow, is not due to some trivial geometric effect but is real. Many factors may play a role in this context, such as magnetic field \citep[see for example the review of][]{Li2014} or binarity \citep{Loinard2002}.

\section{Summary}
In summary, the main original contribution of the present analysis to our understanding of the formation process of L1527 is the presentation of a simple 3D parameterisation of the morphology and kinematics of the gas envelope based solely on the observation of the C$^{18}$O (2-1) line emission from regions that are not dominated by absorption. In the explored range, the model, which uses eight relevant parameters, reproduces observations significantly better than earlier attempts. This was made possible by assuming central symmetry and axisymmetry about the $x$ (west-east) axis and by excluding from the analysis Doppler velocities between 0 and 0.32 \kms and a band of the sky map of $\pm$0.675 arcsec about the $x$ axis. Moreover we have reduced the radial dependences of the temperature and density to a single quantity, the effective emissivity. We have presented arguments showing that such assumptions are indeed reasonable.
The main conclusions of the analysis are:\\
a) {\it{Rotation}}: the model uses successfully trajectories spiralling down toward the disc on paraboloids coaxial with it. We give evidence for the dominance of rotation over the explored range and confirm that the rotation curve is as expected from a transition from an inverse distance law to a Keplerian regime. The rotation velocity decreases from $\sim$1.66 \kms to $\sim$0.34 \kms when $r$ increases from 1 to 5 arcsec. However, the range explored is too distant from the star to allow direct observation of the disc.\\ 
b) {\it{In-fall velocity}}: we measure a mean in-fall velocity of 0.43$\pm$0.10 \kms over the explored range, smaller than free fall velocity, with no evidence for a significant $r$-dependence, in particular no evidence for the sudden jump to $\sim$1.5 \kms required by \citet{Ohashi2014} to describe the absorption-dominated central region.\\
c) {\it{Central mass}}: using simple arguments of energy conservation we evaluate a central mass of 0.23$\pm$0.06 solar masses contained within a distance of 1.5 arcsec from the star, in agreement with earlier estimates obtained from a different range of distances \citep{Tobin2012}. We discuss the systematic uncertainty attached to the fact that the in-fall velocity is increasingly smaller than the free-fall velocity when getting closer to the star.\\ 
d) {\it{Disc plane depression}}: in addition to a strong depression of the in-falling flux about the disc axis, where outflow is known to be present, we observe a similar depression in the neighbourhood of the disc plane, resulting in an $X$ shaped flow directly visible on the data. We argue that it is not an effect of opacity and suggest that it be caused by the freeze-out of CO molecules on cold dust grains occuring preferably near the disc plane.\\
e) {\it{Accretion rate}}: we measure a value of $3.5\pm1.0\,10^{-7}$ \Msold at a distance of 1 arcsec (140 au) from the star, however subject to systematic uncertainties again associated with the deviation from a stationary flow, in relation with the deviation of the in-fall velocity from free fall. This is a factor of nearly 2 smaller than obtained by \citet{Tobin2012} from luminosity arguments. However, we argue that the apparent disagreement should be understood as evidence for the two evaluations actually measuring different quantities.\\
f) {\it{Orientation in space}}: We give an upper limit of $\sim$$22^\circ$ on a possible tilt of the disc plane about the south-north ($y$) axis. A tilt of $\sim$$10^\circ$ about the line of sight ($z$ axis) is observed at large distances ($|y|>$3 arcsec), matching earlier infrared observations \citep{Tobin2008}, but cancels at smaller distances. A same tilt of $10^\circ$ is observed on the map of the mean Doppler velocity.

\section*{Acknowledgements}
This paper makes use of the following ALMA data: ADS/JAO.ALMA\#2012.1.00647.S. ALMA is a partnership of ESO (representing its member states), NSF (USA) and NINS (Japan), together with NRC (Canada), NSC and ASIAA (Taiwan), and KASI (Republic of Korea), in cooperation with the Republic of Chile. The Joint ALMA Observatory is operated by ESO, AUI/NRAO and NAOJ. The data are retrieved from the JVO portal (http://jvo.nao.ac.jp/portal) operated by the NAOJ. 
We are indebted and very grateful to the ALMA partnership, who are making their data available to the public after a one year period of exclusive property, an initiative that means invaluable support and encouragement for Vietnamese astrophysics. We particularly acknowledge friendly support from the staff of the ALMA Helpdesk. We express our deep gratitude to Professors Anne Dutrey and St\'ephane Guilloteau for having introduced us to the physics of star formation and for very helpful comments and suggestions on the the manuscript. We are also deeply grateful to Doctor Pierre Lesaffre for having clarified points related with the dynamics of the in-fall and accretion mechanisms. Financial support is acknowledged from the Vietnam National Satellite Centre (VNSC/VAST), the NAFOSTED funding agency under contract 103.99-2015.39, the World Laboratory, the Odon Vallet Foundation and the Rencontres du Viet Nam.





\begin{thebibliography}{99}

\bibitem[\protect\citeauthoryear{Caselli \& Ceccarelli}{2012}]{Caselli2012}
Caselli, P. \& Ceccarelli, C. 2012, A\&A Rev., 20, 56


\bibitem[\protect\citeauthoryear{Davidson et al.}{2014}]{Davidson2014}
Davidson, J.A., Li, Z.-Y., Hull, C.L.H. et al. 2014, ApJ, 797/2, 74

\bibitem[\protect\citeauthoryear{Fogel et al.}{2011}]{Fogel2011}
Fogel, J.K.J., Bethell, T.J., Bergin, E.A. et al. 2011, ApJ, 726, 29

\bibitem[\protect\citeauthoryear{Gueth et al.}{1997}]{Gueth1997}
Gueth, F., Guilloteau, S., Dutrey, A. \& Bachiller, R. 1997, A\&A, 323, 943G

\bibitem[\protect\citeauthoryear{Guilloteau et al.}{2016}]{Guilloteau2016}
Guilloteau, S., Pi{\'e}tu, V., Chapillon, E. et al. 2016, A\&A, 586, L1

\bibitem[\protect\citeauthoryear{J{\o}rgensen et al.}{2002}]{Jorgensen2002}
J{\o}rgensen, J. K., Sch{\"o}ier, F. L. \& van Dishoeck, E. F. 2002 A\&A 389, 908

\bibitem[\protect\citeauthoryear{J{\o}rgensen et al.}{2015}]{Jorgensen2015}
J{\o}rgensen, J.K., Visser, R., Williams, J.P. and Bergin, E.A. 2015, A\&A, 579, 23

\bibitem[\protect\citeauthoryear{Li et al.}{2014}]{Li2014}
Li, Z.-Y., Banerjee, R., Pudritz, R.E. et al. 2014, {\it The Earliest Stages of Star and Planet Formation: Core Collapse, and the Formation of Disks and Outflows}, arXiv:1401.2219

\bibitem[\protect\citeauthoryear{Loinard et al.}{2002}]{Loinard2002}
Loinard, L., Rodriguez, L.F., D'Alessio, P. et al. 2002, ApJL 581/2, L109

\bibitem[\protect\citeauthoryear{Ohashi et al.}{2014}]{Ohashi2014}
Ohashi, N., Saigo, K., Aso, Y. et al. 2014, ApJ, 796, 131


\bibitem[\protect\citeauthoryear{Pineda et al.}{2012}]{Pineda2012}
Pineda, J. E., Maury, A. J., Fuller, G. A., et al. 2012, A\&A, 544, L7

\bibitem[\protect\citeauthoryear{Sakai et al.}{2014a}]{Sakai2014a}
Sakai, N., Sakai, T., Hirota, T. et al., 2014a, Nature Lett., 507/7490, 78

\bibitem[\protect\citeauthoryear{Sakai et al.}{2014b}]{Sakai2014b}
Sakai, N., Oya, Y., Sakai, T. et al. 2014b, ApJL, 791, L38

\bibitem[\protect\citeauthoryear{Segura-Cox et al.}{2015}]{Segura-Cox2015}
Segura-Cox, D.M., Looney, L.W., Stephens, I.W. 2015, ApJL, 798/1, L2

\bibitem[\protect\citeauthoryear{Tobin et al.}{2008}]{Tobin2008} 
Tobin, J.J., Hartmann, L., Calvet, N. \& D'Alessio, P. 2008, ApJ, 679, 1364

\bibitem[\protect\citeauthoryear{Tobin et al.}{2010}]{Tobin2010} 
Tobin, J.J., Hartmann, L. \& Loinard, L. 2010, ApJ, 722/1, L12

\bibitem[\protect\citeauthoryear{Tobin et al.}{2010b}]{Tobin2010b} 


\bibitem[\protect\citeauthoryear{Tobin et al.}{2012}]{Tobin2012} 
Tobin, J.J., Hartmann, L., Chiang, H.-F. et al. 2012, Nature, 492/7427, 83

\bibitem[\protect\citeauthoryear{Tobin et al.}{2013}]{Tobin2013} 
Tobin, J.J., Hartmann, L., Chiang, H.-F. et al. 2013, ApJ, 771,48


\end{thebibliography}





\bsp	
\label{lastpage}
\end{document}